# Strong coupling expansion for scattering phases in hamiltonian lattice field theories
## II. SU(2) gauge theory in (2 + 1) dimensions

Bernd Dahmen

Deutsches Elektronen-Synchrotron DESY

Notkestrasse 85, D-22603 Hamburg, Germany

**Abstract**

A recently proposed method for a strong coupling analysis of scattering phenomena in hamiltonian lattice field theories is applied to the SU(2) Yang-Mills model in (2 + 1) dimensions. The calculation is performed up to second order in the hopping parameter. All relevant quantities that characterize the collision between the lightest glueballs in the elastic region – cross section, phase shifts, resonance parameters – are determined.

# Introduction

This paper is a continuation of ref. [1] where I have developed a method to obtain strong coupling expansions for scattering quantities in hamiltonian lattice field theories. The keystone of the formalism is the derivation of an effective quantum mechanical system using Bloch's perturbation theory [2]. This system determines the two-particle dynamics of the theory by means of a free hamiltonian and an interaction potential. Thereby the problem of scattering in lattice field theories is reduced to a quantum mechanical problem of scattering at a potential and can be solved using the powerful methods of ordinary quantum mechanics.

In order to improve our understanding of the low-energy regime of elementary particle physics several authors have studied collision phenomena in lattice field theories in recent years [3] – [12]. It is the intent of my approach to supplement these investigations by a discussion within the framework of strong coupling perturbation theory.

In [1] the $(d+1)$ dimensional Ising model was chosen as a test case. The objective here is to investigate the collision of glueballs in non-abelian gauge theories. These models are described by the famous Kogut-Susskind hamiltonian [13], and a strong coupling perturbative analysis was first carried out by Kogut, Sinclair and Susskind in their pioneering work on the SU(3) glueball spectrum [14]. It is characteristic for Yang-Mills theories that the spectrum is rather extensive (to get an idea of the variety of Wilson loops the reader is referred to ref. [15]). In the static limit the excitations are highly degenerate. In particular, one finds a number of single-particle states that have the same static energy as two-glueball configurations. When the next-to-nearest neighbour interaction is turned on, this sector is deformed into a two-channel system describing two identical particles coupled to an additional channel of one-particle states. The reduced hamiltonian is a $2 \times 2$ matrix in the space of the two channels.

As far as the strong coupling analysis of glueball scattering is concerned this observation adds a new aspect compared to the Ising model. In the present paper I will demonstrate to what extent the methods developed with the Ising model must be broadened to be applied to an investigation of non-abelian gauge theories. It will be argued that the effect of the coupling to the single-particle channel can be taken into account by an additional potential in the two-glueball channel.

Since there are a lot more different Wilson loops in three than in two spatial dimensions I decided to discuss the conceptual ideas with the SU(2) gauge theory in $(2 + 1)$ dimensions. The present paper is not self-contained in the sense that the reader is referred to [1] from time to time.

In section 1 the model is introduced. I discuss the static limit and analyse the dynamics of the lightest glueball up to second order in the hopping parameter $\kappa$. Scattering between the lightest glueballs is supposed to take place in the sector of eight-link excitations. Here one distinguishes between the connected Wilson loops and the disconnected configurations. The former are interpreted as single-glueball states whereas the latter represent two distant light particles.

Section 2 is the heart of this work. By an iteration of Bloch's method I show that the eight-link sector decomposes into a number of bound states and an effective two-channel system where the light glueballs are coupled to three heavy particles. As far as the two-glueball channel is concerned an effective Lippmann-Schwinger equation is derived and a series representation for the scattering solutions can be found. The leading order approximation for the scattering amplitude is calculated explicitly.

In section 3 the analysis is completed by the computation of the phase shifts for the different symmetry sectors. The techniques developed with the Ising model require only a slight modification. A resonance is detected in the collision of the light glueballs and discussed in some detail.

I end with some concluding remarks and an outlook. In order to make the paper more readable I have collected some supplementary material as well as a few merely technical items in three appendices.

# 1 SU(2) gauge theory in (2+1) dimensions

## 1.1 Preliminaries

We consider the SU(2) Yang-Mills theory on a rectangular lattice

$$\Gamma = \{\boldsymbol{x} \mid \boldsymbol{x} \in \mathbb{Z}^2,\ -L/2 < x_k \leq L/2,\ k = 1, 2\} \tag{1.1}$$

with linear extent $L$ and lattice spacing $a = 1$. The directed links between neighbouring sites are labeled by a site $\boldsymbol{x}$ and a spatial index $k$ such that the sites $\boldsymbol{x} + \hat{k}$ and $\boldsymbol{x}$ locate the beginning and the end of the link $(\boldsymbol{x}, k)$. By definition, $(\boldsymbol{x}, k)$ and $(\boldsymbol{x} + \hat{k}, -k)$ denote different links although they occupy the same lattice bond. A classical SU(2) lattice gauge field is an assignment of a link variable $U(\boldsymbol{x}, k) \in \text{SU}(2)$ to any bond $(\boldsymbol{x}, k)$ of the lattice $\Gamma$ such that

$$U(\boldsymbol{x}, k)^{-1} = U(\boldsymbol{x} + \hat{k}, -k) . \tag{1.2}$$

On a finite lattice one has to specify boundary conditions. For definiteness we choose them to be periodic, hence

$$U(\boldsymbol{x}, k) = U(\boldsymbol{x} + L\hat{l}, k),\quad l = 1, 2 . \tag{1.3}$$

In the quantum theory the classical field $U(\boldsymbol{x}, k)$ becomes an operator field $\text{U}(\boldsymbol{x}, k)$ acting on some Hilbert space $\mathcal{H}'$ of states. At each link $(\boldsymbol{x}, k)$ of the lattice $\Gamma$ we associate a copy of $L^2(\text{SU}(2))$, the space of square integrable functions on the group manifold (cf. appendix A), and call it $L^2(\boldsymbol{x}, k)$. For a finite lattice the Hilbert space $\mathcal{H}'$ is defined as the tensor product of all $L^2(\boldsymbol{x}, k)$. In this Schrödinger picture the states are complex wave functions $f[U]$ that depend on the basic field variables, i.e. the argument runs over all (periodic) gauge fields $U(\boldsymbol{x}, k)$ on $\Gamma$. The group of (periodic) gauge transformations $g(x)$ on $\Gamma$ acts on the wave functions through

$$f[U] \longrightarrow f[U^g] , \tag{1.4}$$

where $U^g(\boldsymbol{x}, k)$ denotes the gauge transform of the link field

$$U^g(\boldsymbol{x}, k) = g(\boldsymbol{x})\, U(\boldsymbol{x}, k)\, g(\boldsymbol{x} + \hat{k})^{-1} . \tag{1.5}$$

The canonical field operators are identified as multiplication operators on $\mathcal{H}'$

$$[\text{U}(\boldsymbol{x}, k)_{ab} f][U] = U(\boldsymbol{x}, k)_{ab} f[U] . \tag{1.6}$$

In appendix A we introduce the Casimir operator $\text{E}^2$ of the gauge group SU(2). By $\text{E}^2(\boldsymbol{x}, k)$ we denote a copy of $\text{E}^2$ which acts on $L^2(\boldsymbol{x}, k)$ as $\text{E}^2$ and as the identity operator on the other components of the tensor product. Now we are prepared to write down the pure SU(2) lattice gauge theory hamiltonian which was first derived by Kogut and Susskind [13]

$$\mathbb{H}' = (g^2/2) \left[ \sum_{\boldsymbol{x}} \sum_{k=1}^{2} \text{E}^2(\boldsymbol{x}, k) - \kappa \sum_{\boldsymbol{x}} \mathcal{P}(\boldsymbol{x}) \right] . \tag{1.7}$$

Here we have abbreviated the plaquette operator

$$\mathcal{P}(\boldsymbol{x}) = \mathrm{Tr}\left[\mathrm{U}(\boldsymbol{x},1)\,\mathrm{U}(\boldsymbol{x}+\hat{1},2)\,\mathrm{U}(\boldsymbol{x}+\hat{2},1)^{-1}\mathrm{U}(\boldsymbol{x},2)^{-1}\right] . \tag{1.8}$$

and the so-called hopping parameter $\kappa = 2/g^4$. In addition there is the discretized version of Gauss' law with zero external charge. By this constraint the physical Hilbert space is identified with the subspace of $\mathcal{H}'$ that only contains the gauge invariant wave functions. We denote the physical Hilbert space by $\mathcal{H}$.

Let us have a closer look at the structure of the Kogut-Susskind hamiltonian. The first part of (1.7) is a sum over local operators that act non-trivially on one component of the tensor product $\mathcal{H}'$ only

$$\mathbb{H}_0 = \sum_{\boldsymbol{x}} \sum_{k=1}^{2} \mathrm{E}^2(\boldsymbol{x},k) . \tag{1.9}$$

Therefore it is a static term whereas

$$\mathbb{H}_1 = -\sum_{\boldsymbol{x}} \mathcal{P}(\boldsymbol{x}) \tag{1.10}$$

couples the links of the plaquettes and provides a kinetic operator. So we are back with the situation of the Ising model and ready for a perturbative investigation in the hopping parameter $\kappa$. For convenience we introduce

$$\mathbb{H} = (2/g^2)\,\mathbb{H}' = \mathbb{H}_0 + \kappa \mathbb{H}_1 . \tag{1.11}$$

## 1.2 The static limit

The static hamiltonian is a sum of single-link Casimir operators. From the analysis in appendix A we conclude that the eigenfunctions of $\mathbb{H}_0$ are products of representation functions of the gauge group $\mathrm{SU}(2)$. Clearly speaking, for each link $(\boldsymbol{x},k)$ on the lattice $\Gamma$ one chooses a quantum number $n$ that denotes the dimension of the irreducible representation and indices $a,b = 1,\ldots n$. Then one forms the product of the associated representation functions $U_{ab}^n(\boldsymbol{x},k)$. The eigenvalue of this static eigenfunction is the sum of the corresponding one-link eigenvalues $(1/4)(n^2 - 1)$.

The ground state of $\mathbb{H}_0$ is obtained if we set $n = 1$ for all links, i.e. we choose the trivial representation at each link. The corresponding wave function $f[U] = 1$ is gauge invariant, so we are led to interpret this state as the static, physical vacuum and denote it by $|\Omega\rangle^0$:

$$\mathbb{H}_0|\Omega\rangle^0 = 0 , \quad {}^0\langle\Omega\mid\Omega\rangle^0 = 1 . \tag{1.12}$$

The low-lying eigenstates of $\mathbb{H}_0$ have few excited links. Let the operator corresponding to the representation function $U_{ab}^n(\boldsymbol{x},k)$ be denoted by a Roman letter $\mathrm{U}_{ab}^n(\boldsymbol{x},k)$. Then the static state with one excited link $(\boldsymbol{x},k)$ is defined as

$$|\boldsymbol{x},k;n,ab\rangle^0 = N\,\mathrm{U}_{ab}^n(\boldsymbol{x},k)|\Omega\rangle^0 . \tag{1.13}$$

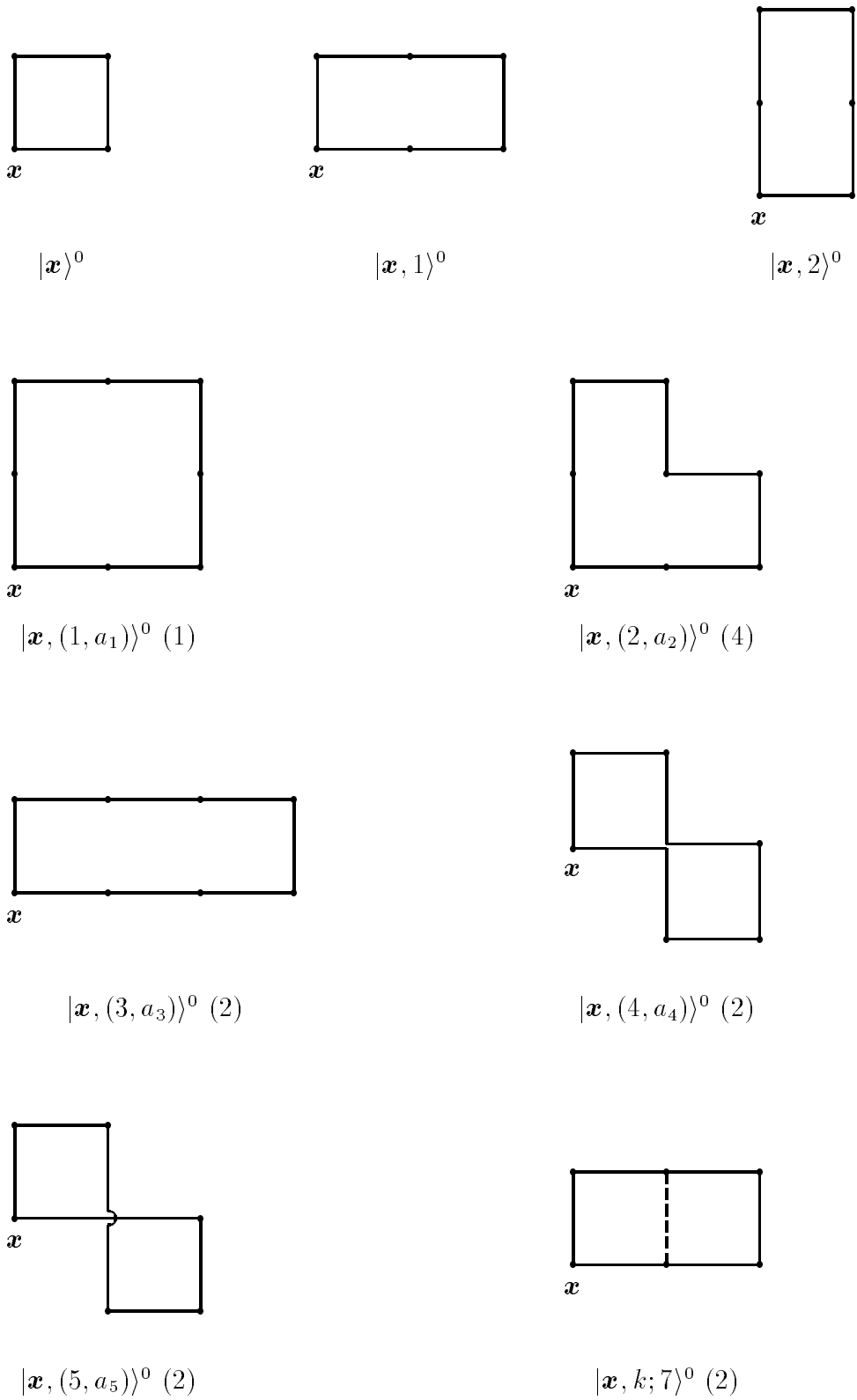

Figure 1.1: Static one-particle excitations.

To determine the value of the normalizing constant $N$ we have to calculate

$$^0\langle \boldsymbol{x}, k; n, ab \mid \boldsymbol{x}, k; n, ab \rangle^0 = N^2 \, {}^0\langle \Omega \mid \mathrm{U}^n_{ab}(\boldsymbol{x}, k)^\dagger \mathrm{U}^n_{ab}(\boldsymbol{x}, k) \mid \Omega \rangle^0 \ . \tag{1.14}$$

The method to evaluate such matrix elements relies on the Clebsch-Gordon series for the product of representation functions and is described in detail in appendix A. Here we obtain $N = \sqrt{n}$. From the definition (1.13) it follows

$$\mathbb{H}_0 |\boldsymbol{x}, k; n, ab\rangle^0 = (1/4)(n^2 - 1)|\boldsymbol{x}, k; n, ab\rangle^0 \ , \tag{1.15}$$

i.e. for each link with quantum number $n > 1$ the colour electric energy is increased by the amount $(1/4)(n^2 - 1)$. We say that a colour electric flux has been created, and the quantum number $n$ is a measure for the strength of this flux.

Within this picture the requirement of gauge invariance for the physical states simply means that the colour electric flux out of any site of the lattice $\Gamma$ is zero. Therefore we have to form appropriate products of the non gauge invariant states (1.13) to generate the physical spectrum of the static hamiltonian $\mathbb{H}_0$. For instance, a gauge invariant excitation is built by taking the trace of $n = 2$ flux elements around one single plaquette

$$|\boldsymbol{x}\rangle^0 \stackrel{\text{def}}{=} \mathcal{P}(\boldsymbol{x})|\Omega\rangle^0 \ , \tag{1.16}$$

where the plaquette operator was defined in eq. (1.8). These states have energy $E_4^{(0)} = 4 \cdot (3/4) = 3$. They satisfy

$$^0\langle \boldsymbol{x}' \mid \boldsymbol{x} \rangle^0 = \delta(\boldsymbol{x}' - \boldsymbol{x}) \ . \tag{1.17}$$

and are interpreted as the lightest static glueballs. Since the trace of SU(2) matrices is real the flux along the plaquette does not have an orientation. The different one-particle states are distinguished by the vector $\boldsymbol{x}$ indicating the position of the plaquette on the lattice.

Heavier glueballs of mass $E_6^{(0)} = 6 \cdot (3/4) = 9/2$ are created from the vacuum by six-link loop operators in the same way as the four-link states (1.16). It is most suggestive and convenient to define them pictorially as in fig. 1.1. The two states $|\boldsymbol{x}, 1\rangle^0$ and $|\boldsymbol{x}, 2\rangle^0$ have the same geometrical shape and are transformed into each other by lattice rotations $\in O(2, \mathbb{Z})$.

The next glueballs in the static spectrum are the loops of length eight with energy $E_8^{(0)} = 6$. There exist five geometrically distinct classes which are invariant under lattice rotations. We adopt the notation

$$|\boldsymbol{x}, (k, a_k)\rangle^0 \ , \tag{1.18}$$

where $k = 1, \ldots, 5$ indicates the class and $a_k$ distinguishes between the different states of the same shape. In fig. 1.1 we depict the five prototypes. The number in brackets denotes the dimensionality of the class.

Since the colour electric energy of the eight-link glueballs is just twice the energy of the four-link states the static eigenspace of energy 6 also contains two-particle excitations

$$|\boldsymbol{x}^1, \boldsymbol{x}^2\rangle^0 = \mathcal{P}(\boldsymbol{x}^1)\mathcal{P}(\boldsymbol{x}^2)|\Omega\rangle^0 \ , \quad |\boldsymbol{x}^1 - \boldsymbol{x}^2| > 1 \ . \tag{1.19}$$

The above restriction is necessary to avoid that the two excited plaquettes overlap on one or more links. In case they do products of representation functions occur which have to be reduced by means of the Clebsch-Gordon series. The resulting states will be closed loops which carry different flux on the (former) common links and represent single-glueball configurations belonging to other static eigenspaces. Since the plaquette operators at different sites commute it follows

$$|\boldsymbol{x}^1, \boldsymbol{x}^2\rangle^0 = |\boldsymbol{x}^2, \boldsymbol{x}^1\rangle^0 \tag{1.20}$$

and the normalization is

$$^0\langle \boldsymbol{y}^1, \boldsymbol{y}^2 \mid \boldsymbol{x}^1, \boldsymbol{x}^2\rangle^0 = {}^0\langle \boldsymbol{y}^1 \mid \boldsymbol{x}^1\rangle^{00}\langle \boldsymbol{y}^2 \mid \boldsymbol{x}^2\rangle^0 + {}^0\langle \boldsymbol{y}^2 \mid \boldsymbol{x}^1\rangle^{00}\langle \boldsymbol{y}^1 \mid \boldsymbol{x}^2\rangle^0 . \tag{1.21}$$

Above the two-particle threshold there are seven-link excitations $|\boldsymbol{x}, k; 7\rangle^0$ whose prototype is depicted in fig. 1.1. Such configurations are created by acting with neighbouring plaquette operators $\mathcal{P}(\boldsymbol{x})$ and $\mathcal{P}(\boldsymbol{x} + \hat{k})$ on the vacuum and taking the $n = 3$ representation on the common link, as indicated by the dashed line. The static colour electric energy is $E_7^{(0)} = 6 \cdot (3/4) + 2 = 13/2$.

This overview illustrates the richness of the (static) spectrum of lattice Yang-Mills theories already in two spatial dimensions. If we go to higher energies we will encounter excited glueball states ($n > 2$), multi-particle configurations and rather long $n = 2$ flux loops.

It is the aim of the present work to investigate the elastic scattering between the lightest glueballs (1.16) by means of a perturbative expansion up to second order in the hopping parameter $\kappa$. As a prerequisite one has to analyse the one-particle dynamics of these excitations. Thereby I repeat the strong coupling technique for gauge theories as described in ref. [14].

## 1.3 Vacuum sector

Since the physical quantities are only defined relative to the ground state of the theory we first have to calculate order by order in $\kappa$ the shift of the static vacuum when the colour magnetic perturbation (1.10) is turned on. From the general formula (2.14) in [1] we obtain the reduced hamiltonian for the vacuum sector

$$H'_0 = \kappa P_0 \mathbb{H}_1 P_0 + \kappa^2 P_0 \mathbb{H}_1 \frac{\mathbb{1} - P_0}{-\mathbb{H}_0} \mathbb{H}_1 P_0 + \mathcal{O}(\kappa^3) , \tag{1.22}$$

where the projection operator is given by $P_0 = |\Omega\rangle^{00}\langle\Omega|$. The action of the plaquette operator $\mathcal{P}(\boldsymbol{x})$ on the vacuum creates an elementary four-link configuration which is orthogonal to the vacuum eigenspace. Consequently there is no first order contribution to $H'_0$.

As for the Ising model we introduce a pictorial description of the processes in strong coupling perturbation theory. In fig. 1.2 the prototype of the non-vanishing matrix elements contributing to the reduced hamiltonian (1.22) in second order is visualized by going from the left to the right. The initial state is the static vacuum without any colour electric flux. The action of the plaquette operator $\mathcal{P}(\boldsymbol{x})$ is represented by a right-arrow and leads to an intermediate four-link configuration of energy 3 at site $\boldsymbol{x}$. Applying $\mathcal{P}(\boldsymbol{x})$ once more the intermediate state is annihilated and we end up with the vacuum as the final state. The value of each such matrix element combined with the energy denominator is $-1/3$. Because there are $L^2$ squares on a two-dimensional periodic lattice of linear extent $L$ the counting factor in $L^2$. One obtains

$$\mathrm{H}'_0 = -\kappa^2 \left(L^2/3\right) \mathrm{P}_0 + \mathcal{O}(\kappa^3) \tag{1.23}$$

as the effective hamiltonian and

$$E_0 = -\kappa^2 \left(L^2/3\right) + \mathcal{O}(\kappa^3) \tag{1.24}$$

for the ground state energy.

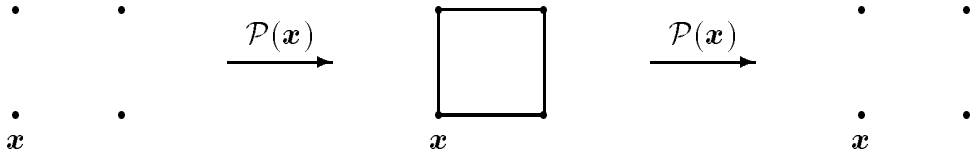

Figure 1.2: Second order vacuum process.

## 1.4 The lightest glueball

The projection operator on the static four-link space $\mathcal{E}_4$ reads

$$\mathrm{P}_4 = \sum_{\boldsymbol{x}} |\boldsymbol{x}\rangle^{00}\langle\boldsymbol{x}| . \tag{1.25}$$

From the general Clebsch-Gordon series (A.21) we know that $2\otimes 2 = 1\oplus 3$, i.e. the product of two fundamental representations does not contain the fundamental representation itself. Therefore the action of the plaquette operator on a four-link state leads out of the static subspace $\mathcal{E}_4$. As for the vacuum sector there is no first order contribution to the reduced hamiltonian.

To second order in perturbation theory one encounters five different types of transitions. The corresponding graphs are collected in fig. 1.3. Column (a) represents the process where the initial state remains undisturbed while a vacuum fluctuation occurs on a distant square. The value of each such matrix element is equal to the vacuum process depicted in fig. 1.2, but the counting factor is reduced by the number of plaquettes that have one link in common with the initial state. In fig. 1.3 (b) the complementary graph is shown where the vacuum fluctuation now has one link in common with the initial state. This link may either be in the trivial representation or in the adjoint representation leading to an intermediate six-link configuration $|\boldsymbol{x},k\rangle^0$ or a seven-link excitation $|\boldsymbol{x},k;7\rangle^0$ (cf. fig. 1.1). The author hopes that the diagrammatic description of the different processes is for the most part self-explanatory such that we must not go through all of the graphs in fig. 1.3. In appendix A all types of elementary matrix elements which are needed to first and second order in perturbation theory throughout this thesis are given explicitly. With these results at hand the evaluation of the different diagrams is straightforward. To be as transparent as possible the values of the matrix elements combined with the energy denominator as well as the counting factors for the processes are also listed in fig. 1.3.

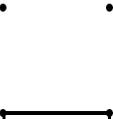

Figure 1.3: Second order contributions to the static four-link sector.
3

We deduce the reduced hamiltonian

$$H'_4 = \left[-(L^2/3) + 29/105\right] \kappa^2 P_4 - (\kappa^2/21) \sum_{\boldsymbol{x}} \sum_{k=1}^{2} \left(|\boldsymbol{x} + \hat{k}\rangle^0 + |\boldsymbol{x} - \hat{k}\rangle^0\right) {}^0\langle\boldsymbol{x}| + \mathcal{O}(\kappa^3) \, . \quad (1.26)$$

The first summand is the result of the graphs fig. 1.3 (a) – (c) where the final state is equal to the initial one. The processes fig. 1.3 (d) – (e) result in a shift of the initial configuration by one plaquette in one of the four spatial directions and lead to the second term of (1.26). Subtracting the vacuum energy shift we define the infinite volume hamiltonian $H_4 = H'_4 - E_0$ and pass to the coordinate space representation

$$H_4 = (3/35) \kappa^2 + (\kappa^2/21) \Delta + \mathcal{O}(\kappa^3) \, , \quad (1.27)$$

where the lattice laplacian was defined in [1, eq. 2.29]. The eigenfunctions are plane waves $\exp(i\boldsymbol{p}\boldsymbol{x})$ and the dispersion relation for the lightest SU(2) glueball in two space dimensions reads

$$E_4(\boldsymbol{p}) = 3 + \kappa^2 \left(3/35 + \hat{\boldsymbol{p}}^2/21\right) + \mathcal{O}(\kappa^3) \, . \quad (1.28)$$

For the physical rest mass and the kinetic mass of the particle we obtain

$$m_\text{r} = 3 + (3/35) \kappa^2 + \mathcal{O}(\kappa^3) \, , \quad m_\text{k}^{-1} = (2/21) \kappa^2 + \mathcal{O}(\kappa^3) \, . \quad (1.29)$$

The reduced hamiltonians for the six- and seven-link sector are obtained in much the same way. In particular, it is straightforward to see that there are no first order contributions either. To second order one finds that the non-vanishing matrix elements (combined with the energy denominator) corresponding to the processes that either shift the initial configuration by one plaquette or transform between the two different states in $\mathcal{E}_6$ and $\mathcal{E}_7$, respectively, add up to zero. Consequently, the heavier particles remain static and degenerate. The mass is

$$m_{\text{r},6} = 9/2 + (8/21) \kappa^2 + \mathcal{O}(\kappa^3) \quad (1.30)$$

for the six-link states and

$$m_{\text{r},7} = 13/2 + (8/21) \kappa^2 + \mathcal{O}(\kappa^3) \quad (1.31)$$

for the seven-link configurations.

# 2 Scattering of glueballs

## 2.1 The eight-link sector as a two-channel system

The static subspace $\mathcal{E}_8$ of energy 6 contains the connected loops of length eight depicted in fig. 1.1 together with pairs of separated four-link loops defined through eq. (1.19). As for the Ising model I restrict the analysis to the center of mass system. Then the two-particle states depend on the relative coordinate $\boldsymbol{r} = \boldsymbol{x}^1 - \boldsymbol{x}^2$ with $|\boldsymbol{r}| > 1$, and the properties (1.20) and (1.21) translate to

$$|\boldsymbol{r}\rangle^0 = |-\boldsymbol{r}\rangle^0 , \quad {}^0\langle \boldsymbol{r}' | \boldsymbol{r}\rangle^0 = \delta(\boldsymbol{r}' - \boldsymbol{r}) + \delta(\boldsymbol{r}' + \boldsymbol{r}) . \tag{2.1}$$

In view of the forthcoming calculations it is advantageous to distinguish between the configurations with a relative distance $|\boldsymbol{r}| > 2$ and those depicted in fig. 2.1. The connected eight-link states are independent of any lattice vector in the center of mass system and are denoted by $|k, a_k\rangle^0$. To fix the notation they are listed in fig. 2.2.

Before we can write down a basis for the linear space $\mathcal{E}_8$ we have to remember a specialty in SU(N) gauge theories, the Mandelstam constraints [16]. These are non-linear identities among loops which intersect or touch at at least one point. Their origin are basic properties of SU(N) matrices. In case of $N = 2$, all Mandelstam constraints can be derived from the following fundamental identity for arbitrary SU(2) matrices $U$ and $V$

$$\operatorname{Tr} UV + \operatorname{Tr} UV^\dagger - \operatorname{Tr} U \operatorname{Tr} V = 0 . \tag{2.2}$$

It is an easy exercise to verify that (2.2) leads to linear relations between certain states in $\mathcal{E}_8$, namely

$$|4,1\rangle^0 + |5,1\rangle^0 - |\hat{1}+\hat{2}\rangle^0 = 0 , \quad |4,2\rangle^0 + |5,2\rangle^0 - |\hat{1}-\hat{2}\rangle^0 = 0 . \tag{2.3}$$

We agree to choose $\{|k, a_k\rangle^0 \mid k = 1\ldots 4\}$ together with the two-loop configurations fig. 2.1 as linear independent states. However, they are not orthogonal, rather

$${}^0\langle 4,1 | \hat{1}+\hat{2}\rangle^0 = {}^0\langle 4,2 | \hat{1}-\hat{2}\rangle^0 = 1/2 . \tag{2.4}$$

In order to obtain an orthonormal basis for the eight-link space we define

$$|4',1\rangle^0 = \left(\sqrt{3}/3\right) \left(2 |4,1\rangle^0 - |\hat{1}+\hat{2}\rangle^0\right) , \quad |4',2\rangle^0 = \left(\sqrt{3}/3\right) \left(2 |4,2\rangle^0 - |\hat{1}-\hat{2}\rangle^0\right) . \tag{2.5}$$

For convenience I introduce a new notation for the states fig. 2.1

$$|5', k\rangle^0 = |2 \cdot \hat{k}\rangle^0 , \quad |6', 1\rangle^0 = |\hat{1}+\hat{2}\rangle^0 , \quad |6', 2\rangle^0 = |\hat{1}-\hat{2}\rangle^0 . \tag{2.6}$$

Below the prime will be dropped whenever there is no danger of confusion.

We define the operators

$$\mathrm{P}'_{\mathrm{sc}} = (1/2) \sum_{|\boldsymbol{r}|>2} |\boldsymbol{r}\rangle^{00}\langle \boldsymbol{r}| , \quad \mathrm{P}'_{\mathrm{b}} = \sum_{k=1}^{6} \sum_{a_k} |k, a_k\rangle^{00}\langle k, a_k| . \tag{2.7}$$

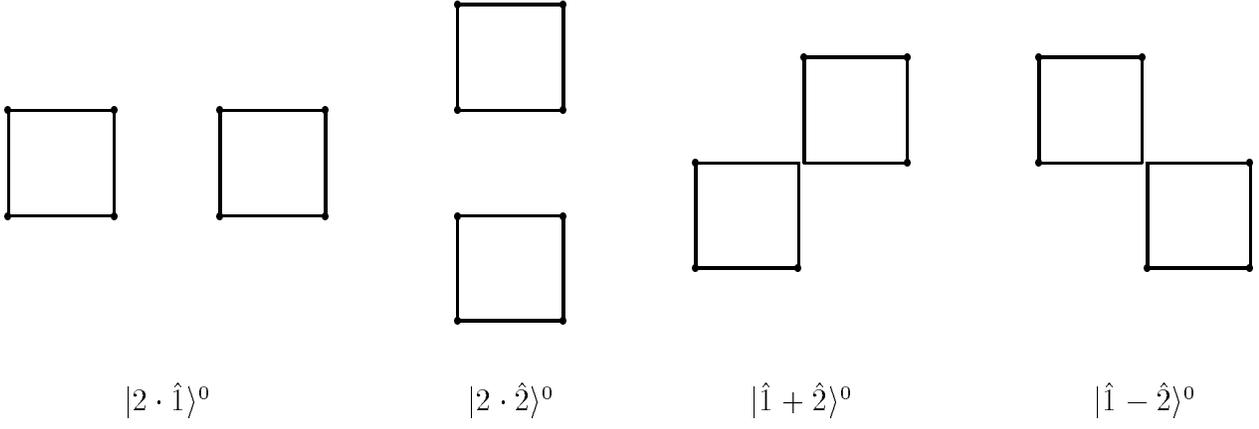

$$|2 \cdot \hat{1}\rangle^0 \qquad |2 \cdot \hat{2}\rangle^0 \qquad |\hat{1}+\hat{2}\rangle^0 \qquad |\hat{1}-\hat{2}\rangle^0$$

Figure 2.1: Two-glueball configurations with a small relative distance $1 < |\boldsymbol{r}| \leq 2$.

The first one, $P'_{sc}$, projects onto the static two-loop configurations with a relative distance larger than 2. When the kinetic hamiltonian $\kappa \mathbb{H}_1$ is turned on, they are deformed into states describing the relative motion of the light glueballs. On the contrary, the states $\{|k, a_k\rangle^0 \mid k = 1, \ldots, 6\}$ stay at rest in the center of mass system and are interpreted as single-particle excitations. Thereby one is led to a natural decomposition of the Hilbert space

$$\mathcal{E}_8 = \mathcal{H}'_{sc} \oplus \mathcal{H}'_{b} \,, \tag{2.8}$$

where $\mathcal{H}'_{sc}$ and $\mathcal{H}'_{b}$ denote the orthogonal subspaces associated to the above projection operators. Accordingly, the states of the system are represented as two-component vectors

$$|\chi\rangle = \begin{pmatrix} |\psi\rangle \\ |\varphi\rangle \end{pmatrix} \,, \quad |\psi\rangle \in \mathcal{H}'_{sc} \,, \quad |\varphi\rangle \in \mathcal{H}'_{b} \,. \tag{2.9}$$

The infinite volume reduced hamiltonian is a $2 \times 2$ matrix

$$H'_8 = \begin{pmatrix} H'_{sc} & V'_{cc} \\ V'^{\dagger}_{cc} & M' \end{pmatrix} \,, \tag{2.10}$$

with

$$H'_{sc} = \kappa P'_{sc} \mathbb{H}_1 P'_{sc} + \kappa^2 P'_{sc} \mathbb{H}_1 \frac{\mathbb{1} - P_8}{6 - \mathbb{H}_0} \mathbb{H}_1 P'_{sc} - E_0 P'_{sc} + \mathcal{O}(\kappa^3) \,, \tag{2.11}$$

$$M' = \kappa P'_{b} \mathbb{H}_1 P'_{b} + \kappa^2 P'_{b} \mathbb{H}_1 \frac{\mathbb{1} - P_8}{6 - \mathbb{H}_0} \mathbb{H}_1 P'_{b} - E_0 P'_{b} + \mathcal{O}(\kappa^3) \,, \tag{2.12}$$

$$V'_{cc} = \kappa P'_{sc} \mathbb{H}_1 P'_{b} + \kappa^2 P'_{sc} \mathbb{H}_1 \frac{\mathbb{1} - P_8}{6 - \mathbb{H}_0} \mathbb{H}_1 P'_{b} + \mathcal{O}(\kappa^3) \,. \tag{2.13}$$

Here $P_8$ represents the projection operator on the entire Hilbert space $\mathcal{E}_8$, hence

$$P_8 = P'_{sc} + P'_{b} \,. \tag{2.14}$$

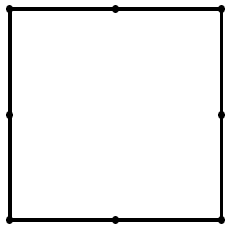
$|1,1\rangle^0$

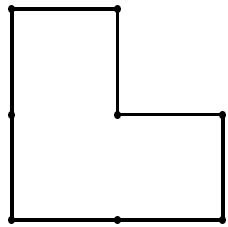
$|2,1\rangle^0$

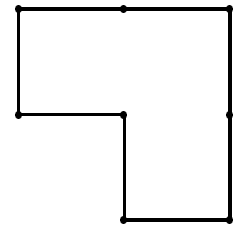
$|2,2\rangle^0$

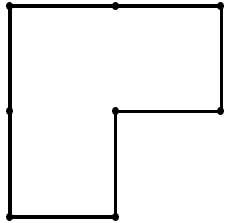
$|2,3\rangle^0$

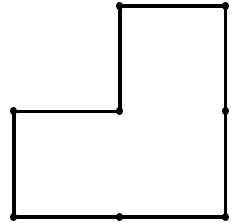
$|2,4\rangle^0$

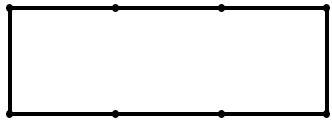
$|3,1\rangle^0$

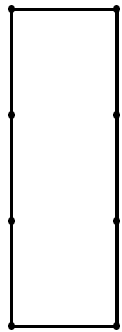
$|3,2\rangle^0$

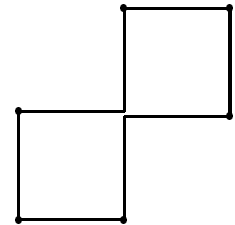
$|4,1\rangle^0$

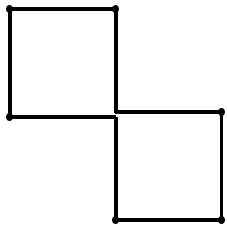
$|4,2\rangle^0$

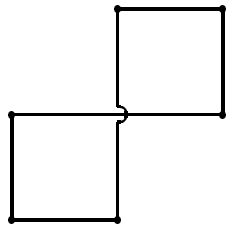
$|5,1\rangle^0$

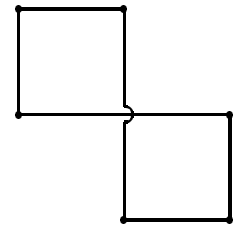
$|5,2\rangle^0$

Figure 2.2: Single-glueball configurations of static energy 6.

To summarize, we are concerned with a two-channel system that describes two identical particles coupled to an additional channel containing 13 single excitations. The hamiltonian $H'_{sc}$ is the operator of the kinetic energy of the two particles possibly supplemented by an interaction potential. In the single-particle channel the hamiltonian $M'$ is just a mass operator. Finally the operator $V'_{cc}$ provides the coupling between the two channels.

## 2.2 The reduced hamiltonian

This section is devoted to the deduction of the reduced hamiltonian (2.10). We start with the two-particle channel.

The first order contributions to $H'_{sc}$ come from processes where the action of the plaquette operator $\mathcal{P}(\boldsymbol{x})$ on a two-glueball initial state leads to a two-glueball final state. From our considerations in section 1 we know that such a transition is not possible.

To second order the perturbation acts on either of the two glueballs as described by the graphs in fig. 1.3, whereas the other four-link loop remains undisturbed. The only thing we have to take care of is the Bose symmetry of the two-particle configurations. This is guaranteed by a four times larger counting factor for the graphs corresponding to fig. 1.3 (b) – (e). One obtains

$$\begin{aligned} H'_{sc} &= \frac{58}{105}\kappa^2 P'_{sc} - \frac{\kappa^2}{21} \sum_{|\boldsymbol{r}|>2} \sum_{k=1}^{2} \left( |\boldsymbol{r} + \hat{k}\rangle^0 + |\boldsymbol{r} - \hat{k}\rangle^0 \right){}^0\langle \boldsymbol{r}| + \frac{2}{21}\kappa^2 \sum_{k=1}^{2} |2\cdot\hat{k}\rangle^{00}\langle 3\cdot\hat{k}| \\ &\quad + \frac{2}{21}\kappa^2 \sum_{k\neq l=1}^{2} \left[ \left(|\hat{k}+\hat{l}\rangle^0 + |2\cdot\hat{k}\rangle^0\right){}^0\langle 2\cdot\hat{k}+\hat{l}| + \left(|\hat{k}-\hat{l}\rangle^0 + |2\cdot\hat{k}\rangle^0\right){}^0\langle 2\cdot\hat{k}-\hat{l}| \right] \\ &\quad + \mathcal{O}(\kappa^3) \, . \end{aligned} \tag{2.15}$$

Since $H'_{sc}$ is defined on the subspace of two-glueball configurations with a relative distance larger than 2 we have to subtract those contributions to the sum over $|\boldsymbol{r}| > 2$ that lead out of $\mathcal{H}'_{sc}$. This is the origin of the additional terms in (2.15).

Next we turn to the mass operator (2.12) in the single-particle channel. Unlike for the two-particle channel first order transitions exist and are shown in fig. 2.3. The plaquette operator creates a four-link excitation on the lower left square having two links in common with the initial state $|2,1\rangle^0$. Since the product of two fundamental representations contains the trivial one we obtain another eight-link configuration, $|4,2\rangle^0$ or $|6,2\rangle^0$, as the final state. Using the methods of appendix A the value of the corresponding matrix elements are computed to be $1/2$ for the graph fig. 2.3 (a) and $1/4$ for the graph fig. 2.3 (b), respectively. Further transitions between states in $\mathcal{H}'_b$ are $|1,1\rangle^0 \to |2,a_2\rangle^0$ belonging to the class fig. 2.3 (a) and $|3,k\rangle^0 \to |5,k\rangle^0$ whose prototype is fig. 2.3 (b). Adding up all contributions the first order reduced hamiltonian in the single-particle channel reads

$$M'_1 = \kappa\left(\mathcal{M}_1 + \mathcal{M}_1^\dagger\right) \, , \tag{2.16}$$

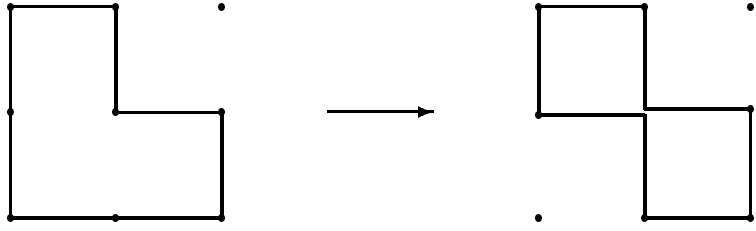

(a)

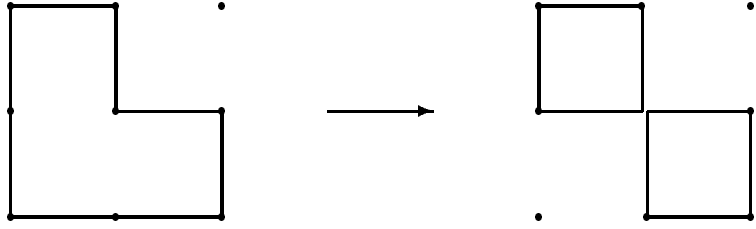

(b)

Figure 2.3: First order transitions.

where we have abbreviated

$$\mathcal{M}_1 \;=\; -\frac{1}{4}\left(2\sum_{k=1}^{4}|2,k\rangle^{00}\langle 1,1| + \sqrt{3}\sum_{k=1}^{2}|2,k\rangle^{00}\langle 4,2| + \sqrt{3}\sum_{k=3}^{4}|2,k\rangle^{00}\langle 4,1|\right.$$
$$\left.+ \sum_{k=1}^{2}|2,k\rangle^{00}\langle 6,2| + \sum_{k=3}^{4}|2,k\rangle^{00}\langle 6,1| + \sum_{k=1}^{2}|3,k\rangle^{00}\langle 5,k|\right)\,.$$

The second order contributions to (2.12) can be grouped together in seven classes, represented by the graphs fig. 1.3 (a) – (c) and fig. 2.4. The values for the corresponding matrix elements are the same for all different processes belonging to the same class whereas the counting factors also depend on the shape of the initial and final states. We get

$$\mathrm{M}_2' = \kappa^2\left(\mathcal{M}_2 + \mathcal{M}_2{}^\dagger\right) + \mathcal{O}(\kappa^3) \tag{2.17}$$

with

$$\mathcal{M}_2 \;=\; \tfrac{17}{168}|1,1\rangle^{00}\langle 1,1| + \tfrac{113}{336}\sum_{k=1}^{4}|2,k\rangle^{00}\langle 2,k| + \tfrac{5057}{29568}\sum_{k=1}^{2}|3,k\rangle^{00}\langle 3,k|$$
$$+\tfrac{35239}{73920}\sum_{k=1}^{2}|4,k\rangle^{00}\langle 4,k| + \tfrac{52037}{147840}\sum_{k=1}^{2}|5,k\rangle^{00}\langle 5,k|$$
$$+\tfrac{31621}{73920}\sum_{k=1}^{2}|6,k\rangle^{00}\langle 6,k| + \tfrac{1}{6}\sum_{k=1}^{2}\sum_{l=3}^{4}|2,l\rangle^{00}\langle 2,k| - \tfrac{1451}{7392}\sqrt{3}\sum_{k=1}^{2}|6,k\rangle^{00}\langle 4,k|\,.$$

| (a) | (b) | (c) | (d) |
|---|---|---|---|
| 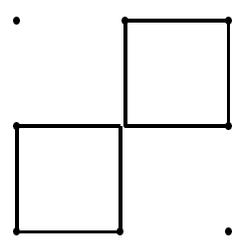 | 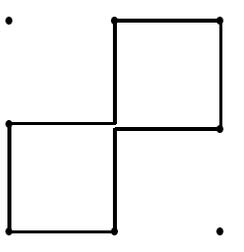 | 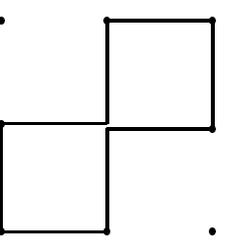 | 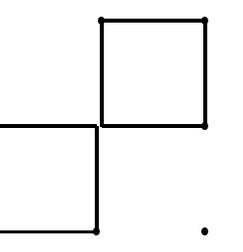 |
| ↑ | ↑ | ↑ | ↑ |
| 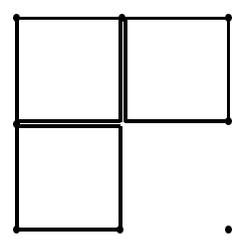 | 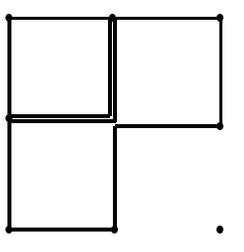 | 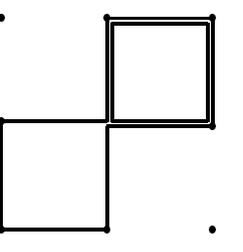 | 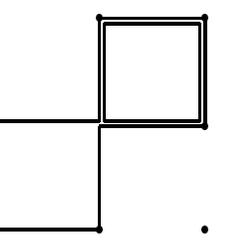 |
| ↑ | ↑ | ↑ | ↑ |
| 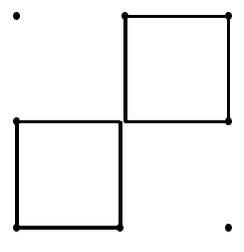 | 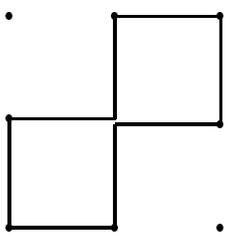 | 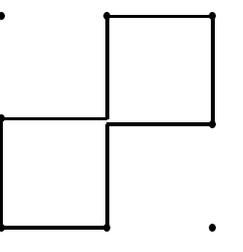 | 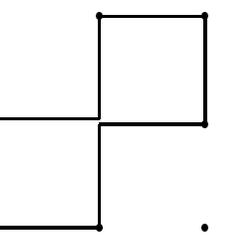 |
| $-195/704$ | $-3/16$ | $-1/15$ | $1/15$ |
| 2 | 2 | 2 | 2 |

Figure 2.4: Second order corrections to the static eight-link sector.

Finally we have to calculate the operator (2.13) that determines the coupling between the two channels. To second order in $\kappa$ the single-particle excitations $\{|k, a_k\rangle^0 \mid k = 4, 5, 6\}$ are coupled to the two-glueball configurations $\{|\boldsymbol{r}\rangle^0 \mid 2 < |\boldsymbol{r}| \leq 3\}$ via the processes fig. 1.3 (d) – (e) that shift one of the excited plaquettes by one square. The result reads

$$\begin{aligned}
V'_{cc} &= -\frac{2}{21}\kappa^2 \sum_{k=1}^{2} |3\cdot\hat{k}\rangle^{0\,0}\langle 5, k| - \frac{2}{21}\kappa^2 \sum_{k\neq l=1}^{2} \left(|2\cdot\hat{k}+\hat{l}\rangle^0 + |2\cdot\hat{k}-\hat{l}\rangle^0\right){}^0\langle 5, k| \\
&\quad + \frac{2}{63}\sqrt{3}\,\kappa^2 \sum_{k\neq l=1}^{2} \Big[|2\cdot\hat{k}+\hat{l}\rangle^0 \left({}^0\langle 4, 1| - \sqrt{3}\,{}^0\langle 6, 1|\right) \\
&\quad\quad + |2\cdot\hat{k}-\hat{l}\rangle^0 \left({}^0\langle 4, 2| - \sqrt{3}\,{}^0\langle 6, 2|\right)\Big] + \mathcal{O}(\kappa^3)\,.
\end{aligned} \qquad (2.18)$$

## 2.3 Diagonalization of the reduced hamiltonian

In the preceding section we have deduced the second order approximation of an operator that determines the dynamics in the eight-link sector $\mathcal{E}_8$. Perturbatively, the eigenvalues and eigenstates of this reduced hamiltonian can be obtained by an iteration of Bloch's method. Clearly speaking, one diagonalizes the first order part exactly and regards the higher order contributions as a perturbation. In case that there are eigenspaces of different energy the problem is further reduced leading to new effective hamiltonians that act on lower dimensional Hilbert spaces.

The first order contribution to $H'_8$ is confined to the single-particle channel and represented by the mass operator (2.16). Among the 13 states $\{|k, a_k\rangle^0 \mid k = 1, \ldots, 6\}$ the degeneracy is partly lifted as $\mathcal{H}'_b$ decomposes in seven orthogonal subspaces. To begin with one finds four one-dimensional and two two-dimensional eigenspaces with masses different from 6. So these excitations decouple from the two-particle channel and represent bound states of the system. Regarding $M'_2$ as a perturbation the second order correction to the various subspaces is determined straightforwardly by means of corresponding reduced hamiltonians. The degeneracy is not reduced any further, and the associated masses are

$$m_5 = 6 - \tfrac{\sqrt{6}}{2}\kappa + \tfrac{71}{105}\kappa^2 + \mathcal{O}(\kappa^3) = 6 - 1.225\,\kappa + 0.676\,\kappa^2 + \mathcal{O}(\kappa^3)\,, \qquad (2.19)$$

$$m_6 = 6 - \tfrac{\sqrt{2}}{2}\kappa + \tfrac{409}{840}\kappa^2 + \mathcal{O}(\kappa^3) = 6 - 0.707\,\kappa + 0.487\,\kappa^2 + \mathcal{O}(\kappa^3)\,, \qquad (2.20)$$

$$m_7 = 6 - \tfrac{1}{4}\kappa + \tfrac{1841}{3520}\kappa^2 + \mathcal{O}(\kappa^3) = 6 - 0.25\,\kappa + 0.523\,\kappa^2 + \mathcal{O}(\kappa^3)\,, \qquad (2.21)$$

$$m_8 = 6 + \tfrac{1}{4}\kappa + \tfrac{1841}{3520}\kappa^2 + \mathcal{O}(\kappa^3) = 6 + 0.25\,\kappa + 0.523\,\kappa^2 + \mathcal{O}(\kappa^3)\,, \qquad (2.22)$$

$$m_9 = 6 + \tfrac{\sqrt{2}}{2}\kappa + \tfrac{409}{840}\kappa^2 + \mathcal{O}(\kappa^3) = 6 + 0.707\,\kappa + 0.487\,\kappa^2 + \mathcal{O}(\kappa^3)\,, \qquad (2.23)$$

$$m_{10} = 6 + \tfrac{\sqrt{6}}{2}\kappa + \tfrac{71}{105}\kappa^2 + \mathcal{O}(\kappa^3) = 6 + 1.225\,\kappa + 0.676\,\kappa^2 + \mathcal{O}(\kappa^3)\,. \qquad (2.24)$$

So for this part of the spectrum the job is done.

In addition, there is a five-dimensional space $\mathcal{H}_b''$ of single-particle excitations that remain degenerate with the two-glueball channel to first order. One obtains a modified two-channel system where the number of additional states coupled to $\mathcal{H}_{sc}'$ is reduced from 13 to 5. In order to construct a modified reduced hamiltonian that acts on the reduced space of the two channels $\mathcal{H}_{sc}' \oplus \mathcal{H}_b''$ and thus determines the scattering of the two four-link glueballs we specify a basis of normalized single-particle states

$$|1'\rangle = (1/2)\left(|4,1\rangle^0 - \sqrt{3}\,|6,1\rangle^0\right) , \tag{2.25}$$

$$|2'\rangle = (1/2)\left(|4,2\rangle^0 - \sqrt{3}\,|6,2\rangle^0\right) , \tag{2.26}$$

$$|3'\rangle = \left(\sqrt{3}/3\right)|1,1\rangle^0 - (1/2)\sum_{k=1}^{2}|4,k\rangle^0 - \left(\sqrt{3}/6\right)\sum_{k=1}^{2}|6,k\rangle^0 , \tag{2.27}$$

$$|4'\rangle = -\left(\sqrt{2}/2\right)\left(|2,1\rangle^0 - |2,2\rangle^0\right) , \tag{2.28}$$

$$|5'\rangle = -\left(\sqrt{2}/2\right)\left(|2,3\rangle^0 - |2,4\rangle^0\right) . \tag{2.29}$$

Introducing the projection operator

$$P_b'' = \sum_{k=1}^{5} |k'\rangle\langle k'| , \tag{2.30}$$

the desired reduced hamiltonian is given by

$$H_8'' = \begin{pmatrix} P_{sc}'\,H_{sc}'\,P_{sc}' & P_{sc}'\,V_{cc}'\,P_b'' \\ P_b''\,V_{cc}'^\dagger\,P_{sc}' & P_b''\,M_2'\,P_b'' \end{pmatrix} . \tag{2.31}$$

From (2.15) it follows that $H_{sc}' = P_{sc}'\,H_{sc}'\,P_{sc}'$, so nothing changes for the two-particle channel. By means of the explicit formula (2.17) the mass operator for the reduced single-particle channel is easily determined

$$P_b''\,M_2'\,P_b'' = \tfrac{2713}{2310}\kappa^2 \sum_{k=1}^{2}|k'\rangle\langle k'| + \tfrac{103}{210}\kappa^2|3'\rangle\langle 3'| - \tfrac{103}{840}\kappa^2 \sum_{k=1}^{2}\left(|k'\rangle\langle 3'| + |3'\rangle\langle k'|\right)$$

$$+ \tfrac{113}{168}\kappa^2 \sum_{k=4}^{5}|k'\rangle\langle k'| + \mathcal{O}(\kappa^3) . \tag{2.32}$$

Having a closer look at the channel coupling potential (2.18) one finds that the first order states (2.25) and (2.26) appear explicitly. Hence, for the coupling between $\mathcal{H}_{sc}'$ and $\mathcal{H}_b''$ one gets

$$P_{sc}'\,V_{cc}'\,P_b'' = \frac{4}{63}\sqrt{3}\,\kappa^2 \sum_{k\neq l=1}^{2}\left(|2\cdot\hat{k}+\hat{l}\rangle^0\langle 1'| + |2\cdot\hat{k}-\hat{l}\rangle^0\langle 2'|\right) . \tag{2.33}$$

As far as the single-glueball channel is concerned the diagonalization of the operator (2.32) leads to four subspaces of different mass

$$m_1 = 6 + \left(\tfrac{641}{770} - \tfrac{\sqrt{12552978}}{9240}\right)\kappa^2 + \mathcal{O}(\kappa^3) = 6 + 0.449\,\kappa^2 + \mathcal{O}(\kappa^3) , \tag{2.34}$$

$$m_2 = 6 + \left(\tfrac{641}{770} + \tfrac{\sqrt{12552978}}{9240}\right)\kappa^2 + \mathcal{O}(\kappa^3) = 6 + 1.216\,\kappa^2 + \mathcal{O}(\kappa^3)\,, \tag{2.35}$$

$$m_3 = 6 + \tfrac{2713}{2310}\kappa^2 + \mathcal{O}(\kappa^3) = 6 + 1.175\,\kappa^2 + \mathcal{O}(\kappa^3)\,, \tag{2.36}$$

$$m_4 = 6 + \tfrac{113}{168}\kappa^2 + \mathcal{O}(\kappa^3) = 6 + 0.673\,\kappa^2 + \mathcal{O}(\kappa^3)\,. \tag{2.37}$$

The first three are of dimension one and can be represented by the normalized states

$$|1\rangle = \beta_1\left(|1'\rangle + |2'\rangle + \alpha_1|3'\rangle\right) + \mathcal{O}(\kappa^2)\,, \tag{2.38}$$

$$|2\rangle = \beta_2\left(|1'\rangle + |2'\rangle + \alpha_2|3'\rangle\right) + \mathcal{O}(\kappa^2)\,, \tag{2.39}$$

$$|3\rangle = -\left(\sqrt{2}/2\right)\left(|1'\rangle - |2'\rangle\right) + \mathcal{O}(\kappa^2)\,. \tag{2.40}$$

The algebraic coefficients have the values

$$\alpha_1 = \tfrac{3160}{1133} + \tfrac{\sqrt{12552978}}{1133} = 5.916\,, \quad \alpha_2 = \tfrac{3160}{1133} - \tfrac{\sqrt{12552978}}{1133} = -0.338\,,$$

and

$$\beta_1 = \left(\alpha_1^2 + 2\right)^{-1/2} = 0.164\,, \quad \beta_2 = \left(\alpha_2^2 + 2\right)^{-1/2} = 0.688\,.$$

The remaining subspace, associated with the mass $m_4$, is of dimension two and spanned by the vectors

$$|4\rangle = |4'\rangle + \mathcal{O}(\kappa^2)\,, \quad |5\rangle = |5'\rangle + \mathcal{O}(\kappa^2)\,. \tag{2.41}$$

From the explicit formula (2.33) it follows that the states (2.41) do not couple to the two-glueball channel $\mathcal{H}'_{\text{sc}}$. They represent two further bound states of the system.

Thus, to second order in the hopping parameter $\kappa$ one is left with an effective quantum mechanical system which describes two identical light glueballs of mass $m_{\text{r}} = 3 + 0.085\,\kappa^2 + \mathcal{O}(\kappa^3)$ coupled to an additional channel that contains three single, heavy particles (2.38) – (2.40) of mass $m_i > 2m_{\text{r}}$. We define the projection operator

$$P_{\text{hp}} = \sum_{k=1}^{3} |k\rangle\langle k| \tag{2.42}$$

and denote the heavy particle channel by $\mathcal{H}_{\text{hp}}$. For notational purposes it is convenient to express the masses of the heavy particles as follows

$$m_i = 6 + \kappa^2\mu_i \qquad \text{for } i = 1, 2, 3\,. \tag{2.43}$$

Then the effective hamiltonian that determines the dynamics in the space of the two channels $\mathcal{H}'_{\text{sc}} \oplus \mathcal{H}_{\text{hp}}$ reads

$$H_8 = \begin{pmatrix} H'_{\text{sc}} & \kappa^2\,V_{\text{cc}} \\ \kappa^2\,V_{\text{cc}}^{\dagger} & \kappa^2\,M \end{pmatrix}\,, \tag{2.44}$$

where the mass operator in the heavy particle channel is given by

$$\kappa^2 \, \mathrm{M} = \mathrm{P}_{\mathrm{hp}} \, \mathrm{P}''_{\mathrm{b}} \, \mathrm{M}'_2 \, \mathrm{P}''_{\mathrm{b}} \, \mathrm{P}_{\mathrm{hp}} = \kappa^2 \sum_{k=1}^{3} \mu_k \, |k\rangle\langle k| \;, \tag{2.45}$$

and the coupling between the two channels is mediated through the potential

$$\kappa^2 \, \mathrm{V}_{\mathrm{cc}} = \mathrm{P}'_{\mathrm{sc}} \, \mathrm{V}'_{\mathrm{cc}} \, \mathrm{P}''_{\mathrm{b}} \, \mathrm{P}_{\mathrm{hp}} = \kappa^2 \, \mathrm{V}^{\mathrm{I}}_{\mathrm{cc}} + \mathcal{O}(\kappa^3) \tag{2.46}$$

with the leading order contribution

$$\begin{aligned}
\mathrm{V}^{\mathrm{I}}_{\mathrm{cc}} &= \frac{4}{63}\sqrt{3} \sum_{k \neq l=1}^{2} \left( |2 \cdot \hat{k} + \hat{l}\rangle^0 + |2 \cdot \hat{k} - \hat{l}\rangle^0 \right) \left( \beta_1 \langle 1| + \beta_2 \langle 2| \right) \\
&\quad - \frac{2}{63}\sqrt{6} \sum_{k \neq l=1}^{2} \left( |2 \cdot \hat{k} + \hat{l}\rangle^0 - |2 \cdot \hat{k} - \hat{l}\rangle^0 \right) \langle 3| \;. 
\end{aligned} \tag{2.47}$$

Note that the coefficients in eq. (2.47) are about 0.08, i.e. the strength of the coupling between the two channels is comparatively weak.

Compared to the previous result (2.10) the effective operator (2.44) is obviously less complicated. So the reduction of the dimensionality really simplifies the problem.

## 2.4 Lippmann-Schwinger equation

As a prerequisite for a determination of the scattering solutions in the coupled two-channel system described above one has to identify the free dynamics. Concerning the two-particle channel we follow the strategy already applied to the Ising model. Any state $|\psi\rangle \in \mathcal{H}'_{\mathrm{sc}}$ has a unique expansion

$$|\psi\rangle = \frac{1}{2} \sum_{\boldsymbol{r}} \psi(\boldsymbol{r}) \, |\boldsymbol{r}\rangle^0 \;, \tag{2.48}$$

where the complex lattice function $\psi(\boldsymbol{r})$ is symmetric and subject to the restrictions $\psi(\boldsymbol{r}) = 0$ for $|\boldsymbol{r}| \leq 2$. The space of these (normalizable) functions is identified as the coordinate representation of the Hilbert space $\mathcal{H}'_{\mathrm{sc}}$. As emphasized in section 2.5 of [1], a wave function that vanishes at certain lattice points near the origin cannot describe the independent relative motion of two particles. Instead we consider the space of unrestricted wave functions $\mathcal{H}_{\mathrm{sc}} \simeq L^2(\mathbb{Z}^2)$. For technical reasons we have also dropped the restriction for the wave functions to be symmetric. At the end of our calculation the true two-glueball sector is restored by projecting on the subspace of symmetric states (cf. the detailed discussion with the Ising model in [1]).

In order to construct a self-adjoint extension of the reduced hamiltonian (2.15) on $\mathcal{H}_{\mathrm{sc}}$ we introduce the operator

$$\mathrm{D}_{\mathrm{nc}} = \mathrm{D}_0 + \sum_{k=1}^{2} \left( \mathrm{D}_{\hat{k}} + \mathrm{D}_{-\hat{k}} + \mathrm{D}_{2 \cdot \hat{k}} + \mathrm{D}_{-2 \cdot \hat{k}} \right) + \mathrm{D}_{\hat{1}+\hat{2}} + \mathrm{D}_{\hat{1}-\hat{2}} + \mathrm{D}_{-\hat{1}+\hat{2}} + \mathrm{D}_{-\hat{1}-\hat{2}} \;. \tag{2.49}$$

It projects on the wave functions that vanish outside a disc of radius 2 around the origin. The extended hamiltonian is defined as

$$H_{sc} = (\mathbb{1} - D_{nc}) H'_{sc} (\mathbb{1} - D_{nc}) + \kappa^2 D(\lambda) , \qquad (2.50)$$

with the local multiplication operator

$$D(\lambda) = \lambda_0 D_0 + \sum_{k=1}^{2} \left[ \lambda_k \left( D_{\hat{k}} + D_{-\hat{k}} \right) + \lambda_{k+2} \left( D_{2 \cdot \hat{k}} + D_{-2 \cdot \hat{k}} \right) \right]$$

$$+ \lambda_5 \left( D_{\hat{1}+\hat{2}} + D_{-\hat{1}-\hat{2}} \right) + \lambda_6 \left( D_{\hat{1}-\hat{2}} + D_{-\hat{1}+\hat{2}} \right)$$

and arbitrary real constants $\lambda_\mu$, $\mu = 0, \ldots, 6$. The explicit result is deduced straightforwardly. One finds

$$H_{sc} = \kappa^2 (T + V_{sc}) , \qquad (2.51)$$

with a kinetic term

$$T = 6/35 + (2/21) \Delta + \mathcal{O}(\kappa) , \qquad (2.52)$$

and a short range potential

$$V_{sc} = -D_{nc} T - T D_{nc} + D_{nc} T D_{nc} + D(\lambda) + \mathcal{O}(\kappa) . \qquad (2.53)$$

As the reader will have expected the kinetic part is just twice the effective hamiltonian (1.27) for the four-link sector. Therefore (2.52) describes the free propagation of the light particles in the two-particle channel. The corresponding (improper) eigenfunctions are the plane waves $\Phi(\boldsymbol{q};\boldsymbol{r}) = \exp(i\boldsymbol{q}\boldsymbol{r})/2\pi$ in the relative motion with momentum $\boldsymbol{q}$. The total kinetic energy of the two glueballs is

$$E_8(\boldsymbol{q}) = 6 + \kappa^2 \varepsilon_8(\boldsymbol{q}) , \quad \varepsilon_8(\boldsymbol{q}) = (6/35) + (2/21) \hat{\boldsymbol{q}}^2 + \mathcal{O}(\kappa) . \qquad (2.54)$$

In the heavy particle channel there is no interaction between the bound states. Concerning the channel coupling potential, eq. (2.46) carries over as it is since $V_{cc}^\dagger$ is trivially extended to an operator on $\mathcal{H}_{sc}$. It follows that the role of the free hamiltonian in the space of the two channels $\mathcal{H}_{sc} \oplus \mathcal{H}_{hp}$ is played by the operator

$$H_0 \stackrel{\text{def}}{=} \kappa^2 \begin{pmatrix} T & 0 \\ 0 & M \end{pmatrix} . \qquad (2.55)$$

Introducing (2.55) the full hamiltonian is split according to

$$H_8 = H_0 + V , \qquad (2.56)$$

where V represents the interaction among the light particles and the coupling between the channels

$$V = \kappa^2 \begin{pmatrix} V_{sc} & V_{cc} \\ V_{cc}^\dagger & 0 \end{pmatrix} . \qquad (2.57)$$

The Møller operators corresponding to this situation are given by

$$\Omega_{\text{in/out}}(H_8, H_0) = \underset{t \to \mp \infty}{\text{s}-\lim} \ e^{iH_8 t} e^{-iH_0 t} P_{\text{sc}} \ . \tag{2.58}$$

The reader is reminded that the wave operators are defined by first projecting on the absolutely continuous subspace of the free hamiltonian (cf. ref. [17], for example). For the present case this is achieved by $P_{\text{sc}}$, the projection operator on $\mathcal{H}_{\text{sc}}$. Since the continuous subspace is spanned by the plane waves the (improper) in- and out-going scattering states are obtained as

$$\begin{pmatrix} \Phi^{\text{in/out}}(q) \\ \varphi^{\text{in/out}}(q) \end{pmatrix} = \Omega_{\text{in/out}} \begin{pmatrix} \Phi(q) \\ 0 \end{pmatrix} \ , \tag{2.59}$$

and it is straightforward to deduce the Lippmann-Schwinger equation. We introduce the Green operators

$$G_{\text{sc}}(q) = \lim_{\rho \searrow 0} \frac{1}{\varepsilon_8(q) - T + i\rho} \ , \quad G_{\text{hp}}(q) = \frac{1}{\varepsilon_8(q) - M} \tag{2.60}$$

that describe the propagation with energy $E_8(q)$ in the two channels. Then the desired integral equation for the in-going scattering solutions of the two-channel system is found to be

$$\begin{pmatrix} \Phi^{\text{in}}(q) \\ \varphi^{\text{in}}(q) \end{pmatrix} = \begin{pmatrix} \Phi(q) \\ 0 \end{pmatrix} + \begin{pmatrix} G_{\text{sc}}(q) & 0 \\ 0 & G_{\text{hp}}(q) \end{pmatrix} \begin{pmatrix} V_{\text{sc}} & V_{\text{cc}} \\ V_{\text{cc}}^\dagger & 0 \end{pmatrix} \begin{pmatrix} \Phi^{\text{in}}(q) \\ \varphi^{\text{in}}(q) \end{pmatrix} \ . \tag{2.61}$$

If one substitute the second component of this equation in the first one can eliminate the heavy particle wave function. Thereby one obtains a modified integral equation for the scattering states of the two-particle channel

$$\Phi^{\text{in}}(q) = \Phi(q) + G_{\text{sc}}(q) \, W(q) \, \Phi^{\text{in}}(q) \ , \tag{2.62}$$

with an energy-dependent potential

$$W(q) = V_{\text{sc}} + V_{\text{cc}} \, G_{\text{hp}}(q) \, V_{\text{cc}}^\dagger \stackrel{\text{def}}{=} V_{\text{sc}} + V_{\text{add}} \ . \tag{2.63}$$

The above result means that, as far as the the two-glueball channel is concerned, the entire effect of the coupling to the heavy particle channel is the replacement of the (original) potential $V_{\text{sc}}$ by the effective potential (2.63). The physical interpretation of $W(q)$ is obvious. The additional term $V_{\text{add}}$ describes a transition from $\mathcal{H}_{\text{sc}}$ to $\mathcal{H}_{\text{hp}}$, propagation there according to $G_{\text{hp}}(q)$ with energy $E_8(q)$, and return to $\mathcal{H}_{\text{sc}}$. As a consequence we must be aware of the occurrence of resonances in the collision of the light glueballs when the total energy exceeds the mass $m_1$ of the lowest lying heavy glueball. Furthermore, since the channel coupling is weak we expect a possible resonance to be sharp.

For the scattering matrix one deduces the familiar result

$$\mathsf{S}(q', q) = (\Phi_{\text{s}}(q'), \Phi_{\text{s}}(q)) - 2\pi i \kappa^2 \, \delta \left( E_8(q') - E_8(q) \right) \left( \Phi_{\text{s}}(q'), W(q) \Phi_{\text{s}}^{\text{in}}(q) \right) \ , \tag{2.64}$$

where the index s indicates that we did not forget to project on the subspace of symmetric wave functions.

## 2.5 Von Neumann series

We are heading for an approximate solution of the Lippmann-Schwinger equation (2.62) in powers of the hopping parameter $\kappa$. The general idea is the same as discussed in section 3.2 of [1]. First we neglect the $\mathcal{O}(\kappa)$ contributions to (2.62) and solve the resulting integral equation exactly. This leads to a modified Lippmann-Schwinger equation which is solved by means of a von Neumann series in the higher order contributions.

The leading order channel coupling (2.46) gives rise to a leading order additional potential

$$V_{\text{add}}^{\text{I}} = V_{\text{cc}}^{\text{I}} \, G_{\text{hp}}(\boldsymbol{q}) \, V_{\text{cc}}^{\text{I}\,\dagger} \, . \tag{2.65}$$

By means of (2.65) the effective potential (2.63) is decomposed as follows

$$W(\boldsymbol{q}) = W^{\text{I}}(\boldsymbol{q}) + \mathcal{O}(\kappa) = V_{\text{sc}}^{\text{I}} + V_{\text{add}}^{\text{I}} + \mathcal{O}(\kappa) \, . \tag{2.66}$$

We substitute (2.66) in the Lippmann-Schwinger equation and obtain

$$\Phi^{\text{in}}(\boldsymbol{q}) = \Phi(\boldsymbol{q}) + G_{\text{sc}}(\boldsymbol{q}) \, W^{\text{I}}(\boldsymbol{q}) \, \Phi^{\text{in}}(\boldsymbol{q}) + \underbrace{G_{\text{sc}}(\boldsymbol{q}) \left[ W(\boldsymbol{q}) - W^{\text{I}}(\boldsymbol{q}) \right]}_{\mathcal{O}(\kappa)} \Phi^{\text{in}}(\boldsymbol{q}) \, . \tag{2.67}$$

Clearly, the third term is of the order $\mathcal{O}(\kappa)$, so we have achieved the desired split of the integral equation for the scattering states in SU(2) gauge theory.

The leading order approximation for the wave function describing the collision of two glueballs is given by the solution of the integral equation

$$\Phi_1^{\text{in}}(\boldsymbol{q}) = \Phi(\boldsymbol{q}) + G_{\text{sc}}(\boldsymbol{q}) \, W^{\text{I}}(\boldsymbol{q}) \, \Phi_1^{\text{in}}(\boldsymbol{q}) \, . \tag{2.68}$$

With this solution at hand the von Neumann series for the scattering states can be derived in complete analogy to the discussion in [1].

## 2.6 Leading order scattering solution

We consider the leading order Lippmann-Schwinger equation in the coordinate space representation

$$\Phi_1^{\text{in}}(\boldsymbol{q};\boldsymbol{r}) = \Phi(\boldsymbol{q};\boldsymbol{r}) + \sum_{\boldsymbol{r}'} G_{\text{sc}}(\boldsymbol{q};\boldsymbol{r},\boldsymbol{r}') \, W^{\text{I}}(\boldsymbol{q}) \, \Phi_1^{\text{in}}(\boldsymbol{q};\boldsymbol{r}') \, . \tag{2.69}$$

The Green function has the integral representation

$$G_{\text{sc}}(\boldsymbol{q};\boldsymbol{r},\boldsymbol{r}') = \int_{\mathcal{B}} \frac{\mathrm{d}^2 p}{(2\pi)^2} \, \frac{\mathrm{e}^{\mathrm{i}\boldsymbol{p}(\boldsymbol{r}-\boldsymbol{r}')}}{\varepsilon_8(\boldsymbol{q}) - \varepsilon_8(\boldsymbol{p}) + \mathrm{i}\rho} \, . \tag{2.70}$$

It fulfils the identity

$$\left[ -T + \varepsilon_8(\boldsymbol{q}) \right] G_{\text{sc}}(\boldsymbol{q};\boldsymbol{r},\boldsymbol{r}') = \delta(\boldsymbol{r} - \boldsymbol{r}') \, , \tag{2.71}$$

and is invariant with respect to planar rotations $\mathrm{O}(2,\mathbb{Z})$ in both coordinate arguments. As a consequence, the integral equation for the symmetrized scattering solutions is obtained if we replace the wave functions in eq. (2.69) by their symmetric counterparts $\Phi_{s,1}^{in}(q;r)$ and $\Phi_s(q;r)$, respectively.

We start the evaluation of the RHS of eq. (2.69) by considering the contributions that are due to the potential $V_{sc}^I$ (cf. eq. (2.53)). The operator $D_{nc}$ projects on those functions in the physical Hilbert space which are zero outside the region $|r| \leq 2$. They are very rapidly decaying and do not belong to the absolutely continuous spectrum. Hence, the symmetric scattering solutions must be orthogonal to these functions, i.e. they vanish at $r = \hat{k}, \hat{k} \pm \hat{l}$. From these considerations we conclude that the operators $TD_{nc}$, $D_{nc}TD_{nc}$ and $D(\lambda)$ do not contribute to the Lippmann-Schwinger equation (2.69).[1] One finds

$$\sum_{r'} G_{sc}(q;r,r') V_{sc}^I \Phi_{s,1}^{in}(q;r') = - \sum_{|\eta| \leq 2} G_{sc}(q;\eta) \xi(q;\eta) . \tag{2.72}$$

The quantities $\xi(q;\eta)$, $|\eta| \leq 2$ result from the action of the operators $D_\eta T$ on $\Phi_{s,1}^{in}(q;r)$. They are sums of values of the symmetric scattering solution at definite coordinate points. Since $\xi(q;\eta) = \xi(q;-\eta)$ there are seven of these unknown functions in eq. (2.72).

The additional potential $V_{add}^I$ is calculated explicitly by inserting the projection operator (2.42) on the heavy particle channel in eq. (2.65)

$$V_{add}^I = V_{cc}^I G_{hp}(q) P_{hp} V_{cc}^{I\,\dagger} = \sum_{k=1}^3 \frac{V_{cc}^I |k\rangle\langle k| V_{cc}^{I\,\dagger}}{\varepsilon_8(q) - \mu_k} . \tag{2.73}$$

The single contributions to this sum are of the type

$$g_{ijk} \frac{|\eta_j\rangle^{00}\langle\eta_i|}{\varepsilon_8(q) - \mu_k} , \quad k = 1,2,3 , \quad 2 < |\eta_i|, |\eta_j| < 3 , \quad g_{ijk} \in \mathbb{R} . \tag{2.74}$$

For sake of completeness I quote the result explicitly

$$\begin{aligned} V_{add}^I &= \sum_{k \neq l=1}^2 \Big[ \Big( g_+ |2 \cdot \hat{k} + \hat{l}\rangle^0 + g_- |2 \cdot \hat{k} - \hat{l}\rangle^0 \Big) \Big( {}^0\langle 2 \cdot \hat{k} + \hat{l}| + {}^0\langle 2 \cdot \hat{l} + \hat{k}| \Big) \\ &\quad + \Big( g_+ |2 \cdot \hat{k} - \hat{l}\rangle^0 + g_- |2 \cdot \hat{k} + \hat{l}\rangle^0 \Big) \Big( {}^0\langle 2 \cdot \hat{k} - \hat{l}| + {}^0\langle 2 \cdot \hat{l} - \hat{k}| \Big) \Big] , \end{aligned} \tag{2.75}$$

where the coefficients $g_+$ and $g_-$ are given by

$$g_\pm = \frac{0.000327}{\varepsilon_8(q) - \mu_1} + \frac{0.00572}{\varepsilon_8(q) - \mu_2} \pm \frac{0.00605}{\varepsilon_8(q) - \mu_3} . \tag{2.76}$$

The rather small numbers in (2.76) are due to the weakness of the coupling between the light and the heavy particles (cf. eq. (2.47)). It follows that the additional potential will only contribute significantly if the energy of the colliding glueballs is near the mass of one of the heavy particles $|1\rangle$, $|2\rangle$ or $|3\rangle$.

---

[1] As a check I also did the calculation without using this a priori information about the behaviour of the scattering solutions near the origin. Like in the case of the Ising model the result corroborates the prediction, i.e. $\Phi_{s,1}^{in}(q,\hat{k}) = \Phi_{s,1}^{in}(q,\hat{k} \pm \hat{l}) = 0$.

In the coordinate space representation of the Hilbert space $\mathcal{H}_{sc}$, (2.74) act as local multiplication operators $D_{\boldsymbol{\eta}}$, $2 < |\boldsymbol{\eta}| < 3$, possibly multiplied with a difference operator. Like $V_{sc}^I$ (cf. eq. (2.53)), the effective potential is short range. For notational convenience we introduce

$$g_{ij}(\boldsymbol{q}) = \sum_{k=1}^{3} \frac{g_{ijk}}{\varepsilon_8(\boldsymbol{q}) - \mu_k} . \tag{2.77}$$

and obtain

$$\sum_{\boldsymbol{r}'} G_{sc}(\boldsymbol{q};\boldsymbol{r},\boldsymbol{r}') V_{add}^I \Phi_{s,1}^{in}(\boldsymbol{q};\boldsymbol{r}') = \sum_{i,j} g_{ij}(\boldsymbol{q}) G_{sc}(\boldsymbol{q};\boldsymbol{r},\boldsymbol{\eta}_i) \Phi_{s,1}^{in}(\boldsymbol{q};\boldsymbol{\eta}_j) . \tag{2.78}$$

The above results eq. (2.72) and eq. (2.78) show that the coordinate dependence of the leading order scattering states is determined by the free solution and the Green function. As unknown quantities there are the sums $\xi(\boldsymbol{q};\boldsymbol{\eta})$, $|\boldsymbol{\eta}| \leq 2$, and the values $\Phi_{s,1}^{in}(\boldsymbol{q};\boldsymbol{\eta}_i)$ of the scattering solution in a disc $2 < |\boldsymbol{\eta}_i| < 3$. Altogether these are 11 complex valued functions which have to be determined in order to solve the leading order Lippmann-Schwinger equation. We adopt the notation

$$\left[\boldsymbol{\eta}_0, \boldsymbol{\eta}_1, \ldots, \boldsymbol{\eta}_{10}\right] = \left[0, \hat{1}, \hat{2}, 2\cdot\hat{1}, 2\cdot\hat{2}, \hat{1}+\hat{2}, \hat{1}-\hat{2}, \hat{1}+2\cdot\hat{2}, \hat{1}-2\cdot\hat{2}, 2\cdot\hat{1}+\hat{2}, 2\cdot\hat{1}-\hat{2}\right] .$$

If one substitutes $\boldsymbol{r} = \boldsymbol{\eta}_i$, $i = 7, \ldots 10$, in eq. (2.69) and acts with $D_{\boldsymbol{\eta}_i} T$, $i = 0, \ldots 6$, on both sides of the Lippmann-Schwinger equation using the identity (2.71), one is left with a set of 11 equations

$$\sum_{j=0}^{6} \mathcal{M}_{ij}\, \xi(\boldsymbol{q};\boldsymbol{\eta}_j) + \sum_{j=7}^{10} \mathcal{M}_{ij}\, \Phi_{s,1}^{in}(\boldsymbol{q};\boldsymbol{\eta}_j) = \varepsilon_8(\boldsymbol{q})\, \Phi_s(\boldsymbol{q};\boldsymbol{\eta}_i) \qquad \text{for } i = 0, \ldots, 7 , \tag{2.79}$$

$$\sum_{j=0}^{6} \mathcal{M}_{ij}\, \xi(\boldsymbol{q};\boldsymbol{\eta}_j) + \sum_{j=7}^{10} \mathcal{M}_{ij}\, \Phi_{s,1}^{in}(\boldsymbol{q};\boldsymbol{\eta}_j) = \Phi_s(\boldsymbol{q};\boldsymbol{\eta}_i) \qquad \text{for } i = 7, \ldots, 10 . \tag{2.80}$$

The elements of the $11 \times 11$ matrix $\mathcal{M}$ consist of sums of Green functions evaluated at definite coordinate arguments, multiplied with the energy-dependent terms $g_{ij}(\boldsymbol{q})$. Due to its complexity an analytical inversion of this object is a formidable task which even the algebraic manipulation program MAPLE [18] is not able to do. However, all quantities appearing in $\mathcal{M}$ are well-behaved functions of the relative momentum of the colliding particles. Hence, for any value of $\boldsymbol{q}$ we are dealing with a complex $11 \times 11$ matrix that MAPLE is able to invert. This provides us with an explicit expression for the unknowns and hence for the leading order scattering solution. Introducing

$$\mathcal{N}_{ij} = \begin{cases} \varepsilon_8(\boldsymbol{q})\, \mathcal{M}_{ij}^{-1} & \text{for } j = 0, \ldots, 6 \\ \mathcal{M}_{ij}^{-1} & \text{for } j = 7, \ldots, 10 , \end{cases} \tag{2.81}$$

the result is

$$\xi(\boldsymbol{q};\boldsymbol{\eta}_i) = \sum_{j=0}^{10} \mathcal{N}_{ij}\, \Phi_s(\boldsymbol{q};\boldsymbol{\eta}_j) \qquad \text{for } i = 0, \ldots, 6 , \tag{2.82}$$

$$\Phi_{s,1}^{in}(\boldsymbol{q};\boldsymbol{\eta}_i) = \sum_{j=0}^{10} \mathcal{N}_{ij}\, \Phi_s(\boldsymbol{q};\boldsymbol{\eta}_j) \qquad \text{for } i = 7, \ldots, 10 . \tag{2.83}$$

## 2.7 Leading order scattering amplitude

In order to obtain the leading order transition matrix one has to calculate the scalar product

$$\mathbb{T}_1(\boldsymbol{q}', \boldsymbol{q}) = \left(\Phi_{\mathrm{s}}(\boldsymbol{q}'), \mathrm{W}^{\mathrm{I}}(\boldsymbol{q})\Phi_{\mathrm{s},1}^{\mathrm{in}}(\boldsymbol{q})\right) = \sum_{\boldsymbol{r}'} \Phi_{\mathrm{s}}(\boldsymbol{q}'; \boldsymbol{r}') \, \mathrm{W}^{\mathrm{I}}(\boldsymbol{q}) \Phi_{\mathrm{s},1}^{\mathrm{in}}(\boldsymbol{q}; \boldsymbol{r}') \ . \qquad (2.84)$$

The action of the local effective potential on the leading order scattering solution has just been worked out. In fact we only have to replace the Green function $G_{\mathrm{sc}}(\boldsymbol{q}; \boldsymbol{r}, \boldsymbol{r}')$ by the symmetric free solution $\Phi_{\mathrm{s}}(\boldsymbol{q}; \boldsymbol{r}')$ in eqs. (2.72) and (2.78). One finds

$$\begin{aligned}\mathbb{T}_1(\boldsymbol{q}', \boldsymbol{q}) &= -\Phi_{\mathrm{s}}(\boldsymbol{q}'; 0)\, \xi(\boldsymbol{q}; 0) - 2 \sum_{i=1}^{6} \Phi_{\mathrm{s}}(\boldsymbol{q}'; \boldsymbol{\eta}_i)\, \xi(\boldsymbol{q}; \boldsymbol{\eta}_i) \\ &\quad + 2 \sum_{i,j=7}^{10} g_{ij}(\boldsymbol{q})\, \Phi_{\mathrm{s}}(\boldsymbol{q}'; \boldsymbol{\eta}_i)\, \Phi_{\mathrm{s},1}^{\mathrm{in}}(\boldsymbol{q}; \boldsymbol{\eta}_j) \ .\end{aligned} \qquad (2.85)$$

In view of the forthcoming calculations it is worthwhile to substitute the results (2.82) and (2.83) for the leading order scattering solution explicitly. We introduce yet another $11 \times 11$ matrix

$$\widetilde{\mathcal{N}}_{ij} = \begin{cases} -\mathcal{N}_{0j} & \\ -2\mathcal{N}_{ij} & \text{for } i = 1, \ldots, 6 \\ 2 \sum_{k=7}^{10} g_{ik}(\boldsymbol{q})\, \mathcal{N}_{kj} & \text{for } i = 7, \ldots, 10 \end{cases} \qquad (2.86)$$

In terms of this quantity the result becomes most concise

$$\mathbb{T}_1(\boldsymbol{q}', \boldsymbol{q}) = \sum_{i,j=0}^{10} \widetilde{\mathcal{N}}_{ij}\, \Phi_{\mathrm{s}}(\boldsymbol{q}'; \boldsymbol{\eta}_i)\, \Phi_{\mathrm{s}}(\boldsymbol{q}; \boldsymbol{\eta}_j) \ . \qquad (2.87)$$

Since both the Green function (2.70) at definite coordinate points and the terms (2.77) only vary with the kinetic energy $E_8(\boldsymbol{q})$ of the colliding glueballs the momentum dependence of the on-shell leading order transition matrix is completely determined by the free solution $\Phi_{\mathrm{s}}(\boldsymbol{q})$. This remarkable observation considerably simplifies the evaluation of (2.87) and hence the diagonalization of the scattering matrix (2.64).

# 3 Phase shifts in lattice gauge theory

The physical quantities that characterize a scattering process are the phase shifts. Making use of the results obtained in the previous sections we can now discuss how the phase shifts for the elastic collision of the lightest SU(2) glueballs in two spatial dimensions are determined. Naturally, many ideas carry over from the detailed discussion of the $(3+1)$ dimensional Ising model in [1]. Therefore we can be rather concise from time to time. The exact eigenvalues of the scattering matrix are written as $\exp(2\mathrm{i}\delta)$. In order to label the phase shifts properly we first consider the low energy regime. At small momentum the rotational symmetry gets restored on the lattice and the phase will behave like in the continuum. They can be identified by their threshold behaviour which is ruled by the angular momentum quantum number $l$ (cf. appendix C.2). For each $l$ one has to distinguish further between the symmetry sectors $\Gamma$ of the discrete symmetry group in two dimensions $\mathrm{O}(2,\mathbb{Z})$ (cf. appendix C.1). So the phase shifts carry the quantum numbers $\delta_l(\Gamma)$.

## 3.1 Threshold behaviour

To begin with we analyse the scattering quantities in the limit where the relative momentum of the colliding particles is very small. The kinetic energy was calculated to second order in the hopping parameter $\kappa$

$$E_8(\boldsymbol{q}) = 2m_\mathrm{r} + \frac{\hat{\boldsymbol{q}}^2}{m_\mathrm{k}} = 2m_\mathrm{r} + \frac{q^2}{m_\mathrm{k}} + \mathcal{O}(q^3) \,, \tag{3.1}$$

where $q = |\boldsymbol{q}|$. We introduce momentum "circular" coordinates

$$q_1 = q\cos(\phi) \,, \quad q_2 = q\sin(\phi) \,,$$

and define energy eigenstates

$$|E_8,\phi\rangle = \left(\frac{1}{q}\frac{\mathrm{d}E_8}{\mathrm{d}q}\right)^{-1/2} \Phi_\mathrm{s}(\boldsymbol{q}) \,, \quad |E_8,\phi\,\mathrm{in}\rangle = \left(\frac{1}{q}\frac{\mathrm{d}E_8}{\mathrm{d}q}\right)^{-1/2} \Phi_\mathrm{s}^\mathrm{in}(\boldsymbol{q}) \,. \tag{3.2}$$

The normalization is chosen such that

$$\langle E_8',\phi' \mid E_8,\phi\rangle = \langle E_8',\phi'\,\mathrm{in} \mid E_8,\phi\,\mathrm{in}\rangle = \delta(E_8' - E_8)\,\delta_\mathrm{s}(\phi' - \phi) \,,$$

where we have abbreviated the symmetrized $\delta$-function

$$\delta_\mathrm{s}(\phi' - \phi) = (1/2)\left[\delta(\phi' - \phi) + \delta(\phi' - \phi - \pi)\right] \,.$$

With respect to these bases of states the scattering matrix assumes the following form

$$\mathsf{S}(E_8',\phi';E_8,\phi) = \mathsf{S}(E_8;\phi',\phi)\,\delta(E_8' - E_8) \,, \tag{3.3}$$

with
$$\mathsf{S}(E_8;\phi',\phi) = \delta_\mathrm{s}(\phi'-\phi) - \pi\mathrm{i}\kappa^2 m_\mathrm{k}\,\mathbb{T}(E_8;\phi',\phi) + \mathcal{O}(q)\;. \tag{3.4}$$

For an analysis of the small momentum behaviour of the leading order transition matrix we cannot use the explicit formula (2.87) since the matrix $\widetilde{\mathcal{N}}$ could not be determined analytically. Instead we go back to eqs. (2.79) and (2.80). The expressions (2.77) as well as the free solution can be expanded in powers of $q$ straightforwardly. As far as the Green function is concerned it is shown in appendix B that

$$G_\mathrm{sc}(\boldsymbol{q};\boldsymbol{\eta}) = \frac{21}{8}\left[\frac{\ln(q)}{2\pi} - \mathrm{i} + b(\boldsymbol{\eta})\right] + \mathcal{O}(q)\;,\quad b(\boldsymbol{\eta})\in\mathbb{R}\;. \tag{3.5}$$

Hence, in order to obtain the threshold behaviour of the scattering solution we can neglect every term that vanishes like $\mathcal{O}(q)$. In particular we replace all cosines by 1 on the RHS of (2.79) and (2.80). Thereby the system of equations considerably simplifies and is easily solved. For small relative momentum the leading order scattering solution has the form

$$\Phi_{\mathrm{s},1}^\mathrm{in}(\boldsymbol{q};\boldsymbol{r}) = \frac{1}{\ln(q)+\pi\,c(\boldsymbol{r})} + \mathcal{O}(q)\;,\quad c(\boldsymbol{r})\in\mathbb{C}\;. \tag{3.6}$$

Substituting in (2.85) we find for the leading order transition matrix

$$\mathbb{T}_1(E_8;\phi',\phi) = \frac{-m_\mathrm{k}^{-1}}{2\pi\,\ln(1.527\,q) - \mathrm{i}\pi^2} + \mathcal{O}(q)\;. \tag{3.7}$$

The above result does not come as a surprise since it is the typical threshold behaviour for finite range potential scattering in two space dimensions. In order to corroborate this statement I discuss a simple quantum mechanical problem in appendix C.2. At low energies the amplitude is isotropic in the center of mass system and the scattering is pure s-wave

$$\delta_0 = \frac{2\pi\,\ln(1.527\,q)}{4\ln^2(1.527\,q)+\pi^2} + \mathcal{O}(q) \quad (\mathrm{mod}\;\pi)\;. \tag{3.8}$$

For the total cross section one finds (cf. appendix C.2)

$$\sigma = \frac{4\pi^2}{q}\,\frac{1}{4\ln^2(1.527\,q)+\pi^2} + \mathcal{O}(q)\;. \tag{3.9}$$

## 3.2 Energy dependence

As in the case of the Ising model we take advantage of the fact that the kinetic energy (up to the order calculated) is a simple polynomial in the absolute value $\widehat{q}=|\widehat{\boldsymbol{q}}|$. Concerning the leading order transition matrix (2.87) it follows that the most complicated contribution, the $11\times 11$ matrix $\widetilde{\mathcal{N}}$, only varies with $\widehat{q}$. Therefore it is sensible to introduce momentum circular variables

$$\widehat{q}_1 = 2\sin(q_1/2) = \widehat{q}\cos(\widehat{\phi})\;,\quad \widehat{q}_2 = 2\sin(q_2/2) = \widehat{q}\sin(\widehat{\phi})\;. \tag{3.10}$$

The associated free and scattering solutions are denoted by $\Phi_{\rm s}(\widehat{q},\widehat{\phi})$ and $\Phi_{\rm s}^{\rm in}(\widehat{q},\widehat{\phi})$, respectively. Their normalization is calculated straightforwardly

$$\left(\Phi_{\rm s}(\widehat{q}',\widehat{\phi}'), \Phi_{\rm s}(\widehat{q},\widehat{\phi})\right) = \left(\Phi_{\rm s}^{\rm in}(\widehat{q}',\widehat{\phi}'), \Phi_{\rm s}^{\rm in}(\widehat{q},\widehat{\phi})\right) = 2m_{\rm k}^{-1} Z(E_8,\widehat{\phi})\,\delta(E_8' - E_8)\,\delta_{\rm s}(\widehat{\phi}' - \widehat{\phi})\,, \quad (3.11)$$

where we have introduced the normalization function

$$Z(E_8,\widehat{\phi}) = \begin{cases} (1/4)\sqrt{16 - 4\widehat{q}^2 + \widehat{q}^4 \sin^2(\widehat{\phi})\cos^2(\widehat{\phi})} & \text{if } -2 < \widehat{q}_k(\widehat{q},\widehat{\phi}) \leq 2 \\ 0 & \text{otherwise} \end{cases} \quad (3.12)$$

The scattering of the two glueballs is invariant with respect to lattice rotations $O(2,\mathbb{Z})$ (this discrete group of planar transformations is discussed in some detail in appendix C.1). Consequently, the (symmetric) eigenstates of the scattering operator are grouped together in multiplets that transform according to the one-dimensional irreducible representations $\Gamma = A_1, A_2, B_1, B_2$ of $O(2,\mathbb{Z})$. This decomposition is made manifest by introducing a new basis for the free and the scattering states

$$\Phi_{\rm s}(\widehat{q},\widehat{\phi}) = \sum_\Gamma \sum_l X_\Gamma^l(\widehat{\phi})\,|E_8;\Gamma,l\rangle\,, \quad \Phi_{\rm s}^{\rm in}(\widehat{q},\widehat{\phi}) = \sum_\Gamma \sum_l X_\Gamma^l(\widehat{\phi})\,|E_8;\Gamma,l\text{ in}\rangle\,. \quad (3.13)$$

The angular functions $X_\Gamma^l$ are defined in appendix C.1. They are a basis for the space of symmetric, infinitely differentiable functions on a unit circle. The index $l$ represents the (continuum) angular momentum quantum number. For each $l$, $X_\Gamma^l$ spans a one-dimensional irreducible subspace for the representation $\Gamma$ of $O(2,\mathbb{Z})$. According to the decomposition (C.2), the colliding particles carry different total angular momentum in the different symmetry sectors. Clearly speaking one has $l = 0, 4, 8, 12, \ldots, \infty$ for $\Gamma = A_1$, $l = 4, 8, 12, \ldots, \infty$ for $\Gamma = A_2$ and $l = 2, 6, 10, \ldots, \infty$ for $\Gamma = B_1, B_2$.

The basis transformations (3.13) reduce the scattering matrix to block diagonal form

$$\mathsf{S}(E_8',\widehat{\phi}'; E_8,\widehat{\phi}) = \delta(E_8' - E_8) \sum_\Gamma \sum_{l,l'} X_\Gamma^{l'}(\widehat{\phi}')^* \mathsf{S}(E_8,\Gamma)_{l'l}\, X_\Gamma^l(\widehat{\phi})\,, \quad (3.14)$$

where

$$\mathsf{S}(E_8,\Gamma)_{l'l} = \mathcal{Z}(E_8,\Gamma)_{l'l} - 2\pi i \kappa^2\, \mathcal{T}(E_8,\Gamma)_{l'l}\,. \quad (3.15)$$

For the normalization matrix $\mathcal{Z}(E_8,\Gamma)$ and the transition matrix $\mathcal{T}(E_8,\Gamma)$ one gets

$$\mathcal{Z}(E_8,\Gamma)_{l'l} = 2m_{\rm k}^{-1} \int d\widehat{\phi}\, Z(E_8,\widehat{\phi}) X_\Gamma^{l'}(\widehat{\phi}) X_\Gamma^l(\widehat{\phi})^*\,, \quad (3.16)$$

$$\mathcal{T}(E_8,\Gamma)_{l'l} = \int d\widehat{\phi}' \int d\widehat{\phi}\, X_\Gamma^{l'}(\widehat{\phi}')\, \mathbb{T}(E_8;\widehat{\phi}',\widehat{\phi})\, X_\Gamma^l(\widehat{\phi})^*\,. \quad (3.17)$$

In order to determine the phase shifts it remains to diagonalize the modified scattering matrix

$$\mathsf{S}'(E_8,\Gamma) = \mathcal{Z}(E_8,\Gamma)^{-1}\, \mathsf{S}(E_8,\Gamma) = \mathbb{1} - 2\pi i \kappa^2\, \mathcal{Z}(E_8,\Gamma)^{-1}\, \mathcal{T}(E_8,\Gamma)\,. \quad (3.18)$$

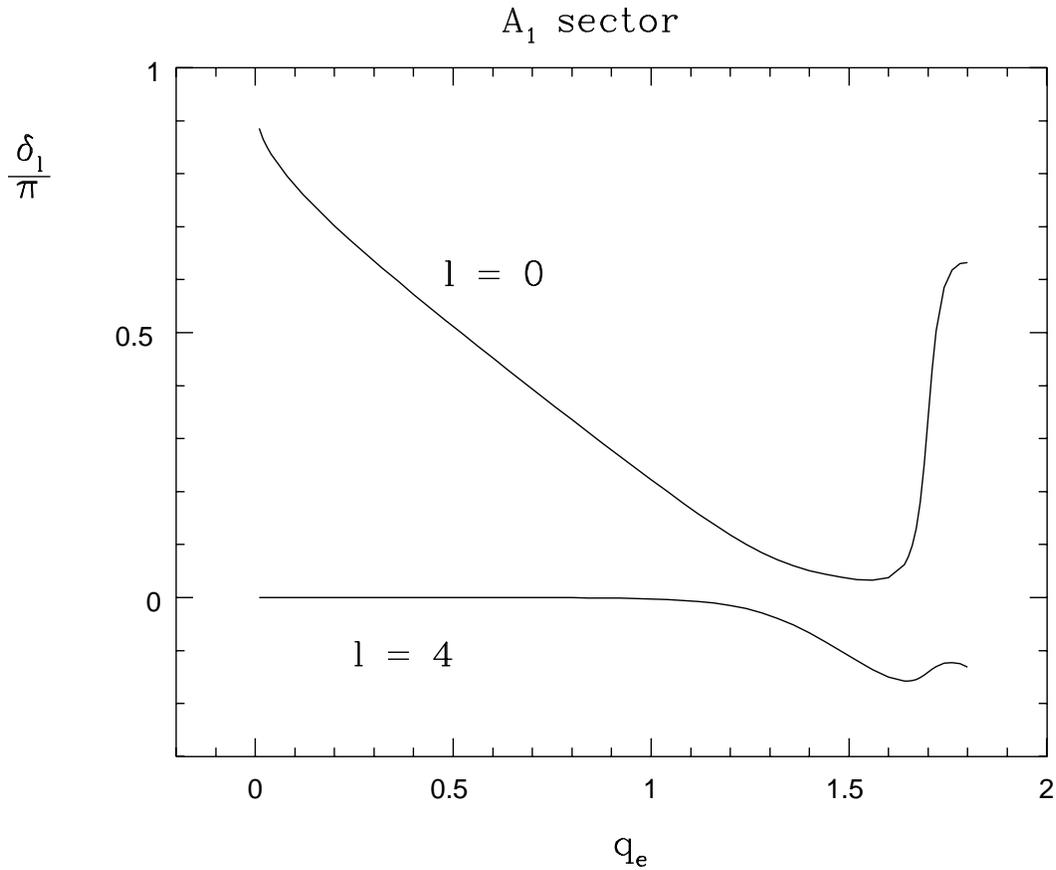

Figure 3.1: The scattering phases in the $A_1$ sector as a function of the momentum $q_{\rm e}$.

As long as $\hat{q} < 2$ the integration over the angular variable in (3.16) is unrestricted (cf. the definition of the normalization function (3.12)), and one can use the orthonormality of the angular functions $X_\Gamma^l(\hat{\phi})$ to invert the normalization matrix

$$\mathcal{Z}(E_8, \Gamma)_{l'l}^{-1} = (1/2) m_{\rm k} \int {\rm d}\hat{\phi}\, Z(E_8, \hat{\phi})^{-1} X_\Gamma^{l'}(\hat{\phi})\, X_\Gamma^l(\hat{\phi})^* \ . \tag{3.19}$$

The inverse normalization function has a series representation in powers of the momentum square $\hat{q}^2$. The coefficients are trigonometric functions of the angular variable $\hat{\phi}$ and can be expressed through exponentials. Obviously, the same is true for $X_\Gamma^l(\hat{\phi})$. It follows that each matrix element (3.19) is a sum of one-dimensional integrals over a product of three exponential functions. Those integrals are zero unless the exponents add up to zero.

Concerning the leading order transition matrix $\mathbb{T}_1(E_8; \phi', \phi)$ the angular dependence is through the free solutions only (cf. eq. (2.87)). To be explicit, if $(\eta_1, \eta_2)$ denote the two components of the lattice vector $\boldsymbol{\eta}$, one has

$$\Phi_{\rm s}(\boldsymbol{q}; \boldsymbol{\eta}) = (2\pi)^{-1} \cos(\eta_1 q_1 + \eta_2 q_2) = (2\pi)^{-1}[\cos(\eta_1 q_1)\cos(\eta_2 q_2) - \sin(\eta_1 q_1)\sin(\eta_2 q_2)] \ .$$

Now both $\cos(\eta_k q_k)$ and $\sin(\eta_k q_k)$ can be decomposed into a sum whose contributions are of the type

$$\cos^\mu(q_k) \sin^\nu(q_k) \ , \qquad \mu, \nu \leq |\eta_k| \ , \quad \mu, \nu \in \mathbb{N}_0 \ .$$

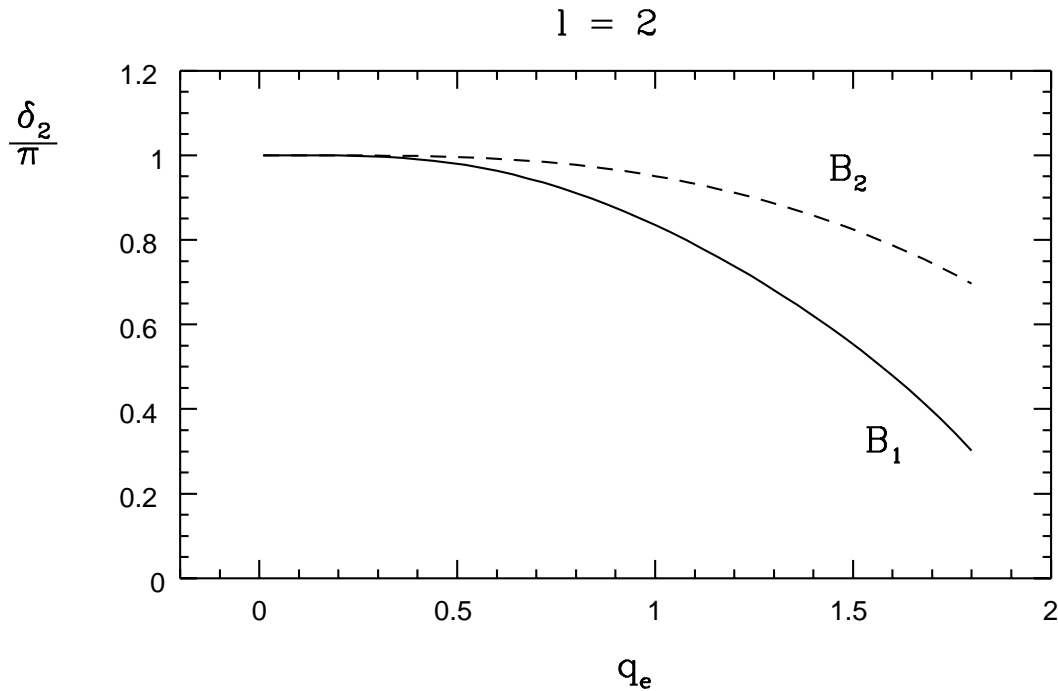

Figure 3.2: Partial d-wave phase shifts occur in the $B_1$ and $B_2$ sector.

From the coordinate transformation (3.10) it follows that

$$\cos(q_k) = 1 - \hat{q}_k^2/2 \,, \quad \sin(q_k) = (\hat{q}_k/2)\sqrt{4 - \hat{q}_k^2} \,.$$

Hence $\mathbb{T}_1(E_8; \phi', \phi)$ has a series representation in powers of $\hat{q}$ as well, and the coefficients are trigonometric functions of the primed and unprimed angular degrees of freedom. These terms can be expressed as sums of representation functions $X_\Gamma^l(\hat{\phi}')$ and $X_\Gamma^l(\hat{\phi})$ such that the integrals (3.17) can be evaluated algebraically. If we tolerate some small error we can truncate the series for (3.17) and (3.19) at some order $\mathcal{O}(\hat{q}^N)$ (for $\hat{q} = 1.8$ we need $N \simeq 50$ to guarantee an error of less than 1%). As a consequence these matrices become finite dimensional.

Now the scattering matrix (3.18) can be diagonalized straightforwardly in each symmetry sector $\Gamma$. The associated (non-vanishing) phase shifts are plotted in figs. 3.1 and 3.2 as a function of the on-shell momentum

$$q_e = \sqrt{m_k(E_8 - 2m_r)} = \hat{q} + \mathcal{O}(\kappa^3) \qquad (3.20)$$

(in [1] the corresponding quantity was named $q_\star$). These results represent the leading order approximation of the scattering phases characterizing the elastic collision between the lightest SU(2) glueballs in two space dimensions. We find effects in the partial s-wave, d-wave and even g-wave, on the proviso that there may be significant lattice artefacts in the region $q_e > 1$. The observed behaviour indicates a strong attractive interaction between the two glueballs.

## 3.3 The effect of the heavy particle channel

In the $A_1$ sector, around $q_e \simeq 1.7$, the $l = 0$ scattering phase shows a behaviour that is typical in presence of a sharp resonance. Indeed, at $q_e = 1.707$ the two-glueball energy $E_8(\boldsymbol{q})$ is equal to the mass $m_1$ of the lowest lying excitation in the heavy particle channel. So we are in a region where the additional potential (2.75) is relevant. Note that the latter only contributes to the $A_1$ sector since the heavy particles (2.38 – 2.40) have the appropriate quantum number.

Instead of determining the resonance parameters from the data fig. (3.1) one may try to investigate the behaviour of the system in the vicinity of the resonance by a suitable approximation. If $E_8(\boldsymbol{q}) \simeq m_1$ the dominant contribution to the Lippmann-Schwinger equation comes from the additional potential (2.75). Moreover, we can approximate

$$g_+ \simeq g_- \simeq \frac{0.000327}{\varepsilon_8(\boldsymbol{q}) - \mu_1} \stackrel{\text{def}}{=} \frac{\lambda^2}{\varepsilon_8(\boldsymbol{q}) - \mu_1}$$

and obtain

$$V^{\text{I}}_{\text{add}} \simeq \frac{\lambda^2}{\varepsilon_8(\boldsymbol{q}) - \mu_1} \sum_{k \neq l=1}^{2} \sum_{k' \neq l'=1}^{2} \left( |2 \cdot \hat{k} + \hat{l}\rangle^0 + |2 \cdot \hat{k} - \hat{l}\rangle^0 \right)$$
$$\times \left( {}^0\langle 2 \cdot \hat{k}' + \hat{l}'| + {}^0\langle 2 \cdot \hat{k}' - \hat{l}'| \right) . \qquad (3.21)$$

It is advantageous to consider this quantity as the effective potential of a simplified two-channel resonance model (cf. ref. [19]). To this aim we introduce the hamiltonian

$$H_{\text{res}} = \begin{pmatrix} T & V_{\text{res}} \\ V_{\text{res}}^\dagger & \mu_1 \end{pmatrix} \qquad (3.22)$$

that acts on the Hilbert space built by the two-glueball configurations $\mathcal{H}_{\text{sc}}$ and the heavy state $|1\rangle$. The coupling between the two channels is represented by the potential

$$V_{\text{res}} = \lambda \sum_{k \neq l=1}^{2} \left( |2 \cdot \hat{k} + \hat{l}\rangle^0 + |2 \cdot \hat{k} - \hat{l}\rangle^0 \right) \langle 1| . \qquad (3.23)$$

Within this system the collision of the light glueballs is determined by the Lippmann-Schwinger equation

$$\begin{pmatrix} \Phi_{\text{s}}^{\text{in}}(\boldsymbol{q}) \\ \varphi^{\text{in}}(\boldsymbol{q}) \end{pmatrix} = \begin{pmatrix} \Phi_{\text{s}}(\boldsymbol{q}) \\ 0 \end{pmatrix} + \begin{pmatrix} G_{\text{sc}}(\boldsymbol{q}) & 0 \\ 0 & [\varepsilon_8(\boldsymbol{q}) - \mu_1]^{-1} \end{pmatrix} \begin{pmatrix} 0 & V_{\text{res}} \\ V_{\text{res}}^\dagger & 0 \end{pmatrix} \begin{pmatrix} \Phi_{\text{s}}^{\text{in}}(\boldsymbol{q}) \\ \varphi^{\text{in}}(\boldsymbol{q}) \end{pmatrix} . \qquad (3.24)$$

As discussed above (cf. eq. (2.62)) one could easily derive a modified equation for the two-particle channel with an effective potential

$$W_{\text{res}}(\boldsymbol{q}) = \frac{V_{\text{res}} V_{\text{res}}^\dagger}{\varepsilon_8(\boldsymbol{q}) - \mu_1}$$

that is nothing but the approximate potential (3.21) we started from.

However, we will step along the complementary path here. We will not eliminate the heavy particle wave function but deduce an explicit representation for it. Then, by means of the Lippmann-Schwinger equation (3.24) one finds for the on-shell transition matrix

$$\mathbb{T}(\bm{q}',\bm{q}) = \delta\left[\varepsilon_8(\bm{q}') - \varepsilon_8(\bm{q})\right] \left(\Phi_\mathrm{s}(\bm{q}), \mathrm{V}_\mathrm{res}\,\varphi^\mathrm{in}(\bm{q})\right) \; . \tag{3.25}$$

Acting with $\mathrm{V}_\mathrm{res}^\dagger$ on the first component of eq. (3.24), substituting the second component and inserting the result in the second component again, it follows

$$\varphi^\mathrm{in}(\bm{q}) = \frac{\mathrm{V}_\mathrm{res}\,\Phi_\mathrm{s}(\bm{q})}{\varepsilon_8(\bm{q}) - \mu_1 - \mathrm{V}_\mathrm{res}^\dagger\,G_\mathrm{sc}\,\mathrm{V}_\mathrm{res}} \; . \tag{3.26}$$

Now we substitute this expression for the heavy particle wave function in eq. (3.25). The result is the famous Breit-Wigner formula describing the rapid energy dependence of the scattering amplitude near a resonance (cf. ref. [20], for example)

$$\mathbb{T}(\bm{q}',\bm{q}) = \frac{\tfrac{1}{2}\Gamma(\bm{q}',\bm{q})}{\varepsilon_8(\bm{q}) - \varepsilon_\mathrm{res}(\bm{q}) + \mathrm{i}\tfrac{1}{2}\Gamma(\bm{q})}\, \delta\left[\varepsilon_8(\bm{q}') - \varepsilon_8(\bm{q})\right] \; . \tag{3.27}$$

The quantities in the denominator are given by

$$\varepsilon_\mathrm{res}(\bm{q}) = \mu_1 + \lambda^2 \sum_{i,j=7}^{10} \mathrm{Re}\left[G_\mathrm{sc}(\bm{q};\bm{\eta}_i,\bm{\eta}_j) + G_\mathrm{sc}(\bm{q};\bm{\eta}_i,-\bm{\eta}_j)\right] \; ,$$

$$\Gamma(\bm{q}) = -2\lambda^2 \sum_{i,j=7}^{10} \mathrm{Im}\left[G_\mathrm{sc}(\bm{q};\bm{\eta}_i,\bm{\eta}_j) + G_\mathrm{sc}(\bm{q};\bm{\eta}_i,-\bm{\eta}_j)\right]$$

$$= 4\pi\lambda^2 \sum_{i,j=7}^{10} \int_\mathcal{B} \mathrm{d}^2 p\,\delta\left(\varepsilon_8(\bm{q}) - \varepsilon_8(\bm{p})\right)\,\Phi_\mathrm{s}(\bm{q};\bm{\eta}_i)\,\Phi_\mathrm{s}(\bm{p};\bm{\eta}_j) \; ,$$

whereas the numerator reads

$$\Gamma(\bm{q}',\bm{q}) = 4\lambda^2 \sum_{i,j=7}^{10} \Phi_\mathrm{s}(\bm{q}';\bm{\eta}_i)\,\Phi_\mathrm{s}(\bm{q};\bm{\eta}_j)$$

$$= (4/\pi)\lambda^2\left(f_0\,X^0_{A_1}(\widehat{\phi}') + f_4\,X^4_{A_1}(\widehat{\phi}')\right)\left(f_0\,X^0_{A_1}(\widehat{\phi}) + f_4\,X^4_{A_1}(\widehat{\phi})\right)$$

with

$$f_0 = \sqrt{2}\left(2 - \tfrac{5}{2}\widehat{q}^2 + \tfrac{5}{8}\widehat{q}^4 - \tfrac{1}{32}\widehat{q}^6\right) \; , \quad f_4 = -\tfrac{1}{8}\widehat{q}^4 + \tfrac{1}{32}\widehat{q}^6 \; .$$

In order to determine the parameters of the resonances it remains to locate the pole in eq. (3.27). Obviously, since all quantities are known explicitly, this is easily done, and the wanted momentum value is $q_\mathrm{e} = 1.712$. So the resonance in the collision of the light glueballs occurs at the energy

$$E_\mathrm{res} = 6 + \kappa^2\,\varepsilon_\mathrm{res}(1.712) = 6 + 0.451\,\kappa^2 + \mathcal{O}(\kappa^3) \tag{3.28}$$

and has the width

$$\Gamma_\mathrm{res} = \kappa^2\,\Gamma(1.712) = 0.0148\,\kappa^2 + \mathcal{O}(\kappa^3) \; . \tag{3.29}$$

# Concluding remarks and outlook

The present paper demonstrates that a systematic strong coupling analysis of glueball scattering is feasible. For the case of SU(2) Yang-Mills theory in (2 + 1) dimensions the phase shifts describing the elastic collision of the lightest glueballs could be calculated explicitly to leading order. A resonance was detected whose parameters could be determined accurately.

Further investigations might go in two directions.

The first is to extend the present work to higher orders in the inverse coupling. This may be worthwhile since only the leading order approximation of the scattering quantities is known so far. Thereby one would achieve further insight into the nature of glueball scattering and could clarify the role of the resonance. I do not expect to meet any severe difficulties, although the diagonalization of the scattering matrix will be more complicated since the momentum dependence of the two-particle energy will probably be less simple to higher orders in the hopping parameter. From a technical point of view it would be desirable to develop an algorithm that generates the higher order reduced hamiltonians by computer.

The second and maybe more interesting possibility is the investigation of lattice models that are of direct physical relevance. This would mean to go from two to three space dimensions (from the point of view of strong coupling perturbation theory the differences between SU(2) and SU(3) are negligible). The number of geometrically distinct Wilson loops of length eight increases from 5 to 15, and the different classes contain up to 48 states. In addition, the four-link excitations receive a further (spin) degree of freedom since the plaquette may be oriented in either of the spatial directions. Clearly, the eight-link sector, which is the staring point for the derivation of an effective two-glueball hamiltonian, is much more involved than in two space dimensions. However, I would like to emphasize that the resulting difficulties are not due to an unsuitable method but reflect the complexity of the theory itself.

From the conceptual point of view all aspects are already present in the two-dimensional model. Therefore the procedure carries over almost verbatim and the problem can be tackled following the way described in great detail in the present work. In particular, one can be sure that the spectrum decouples to higher orders of the hopping parameter such that the effective two-particle system simplifies. Still it must be expected that there are a number of resonances in the collision of two glueballs. Such a strong coupling analysis would help to clarify an early result of Münster [25] who found that the glueball interaction strength is extremely large. A possible explanation could be the existence of a resonance near threshold.

In principle there is no obstacle to apply the concepts developed in [1] and this work for a qualitative investigation of the low-energy scattering of mesons and baryons. One interesting question which seems to be within reach of a strong coupling analysis is whether, in the presence of light quarks, the decay of glueballs into pions is really suppressed as predicted by the phenomenological Zweig rule.

# Acknowledgements

I take the opportunity to thank Martin Lüscher for many stimulating discussions and a critical reading of the manuscript. Financial support by Studienstiftung des deutschen Volkes is gratefully acknowledged.

# Appendix A

## A.1 Elementary properties of the group $SU(2)$

The Lie group $SU(2)$ consists of all unitary $(2 \times 2)$ matrices with determinant equal to 1. The corresponding Lie algebra $su(2)$ can be identified with the space of all complex $(2 \times 2)$-matrices $X$ which satisfy

$$X = -X^\dagger , \quad \text{Tr}\, X = 0 . \tag{A.1}$$

The matrix exponential

$$\exp(X) = \mathbb{1} + \sum_{j=1}^{\infty} \frac{X^j}{k!} \tag{A.2}$$

defines a map from $su(2)$ onto $SU(2)$. The Lie algebra is a real vector space of dimension 3. We choose a basis of hermitean matrices $T_a$, $a = 1, 2, 3$ such that

$$\text{Tr}\, T_a T_b = (1/2)\, \delta_{ab} . \tag{A.3}$$

They satisfy

$$[T_a, T_b] = i\varepsilon_{abc} T_c . \tag{A.4}$$

where $\varepsilon_{abc}$ is the totally antisymmetric symbol on three indices. To be explicit, one often uses the familiar Pauli matrices

$$T_a = \sigma_a/2 . \tag{A.5}$$

With respect to such a basis any element $X \in su(2)$ can be written as

$$X = iX_a T_a , \quad X_a \in \mathbb{R} . \tag{A.6}$$

For complex functions $f(U)$ on $SU(2)$ a natural inner product can be introduced through

$$(f, g) = \int dU f(U)^* g(U) . \tag{A.7}$$

Here $dU$ denotes the invariant group (Haar) measure which satisfies $d(gUh^{-1}) = dU$ for all $g, h \in SU(2)$. The associated square integrable functions represent the Hilbert space $L^2(SU(2))$. For any group element $g \in SU(2)$ one may define an operator $R(g)$ that translates the arguments $U$ of the wave functions

$$[R(g)f](U) = f(g^{-1}U) . \tag{A.8}$$

Thereby a representation of the group $SU(2)$ on the linear space $L^2(SU(2))$ is introduced. Since the scalar product (A.7) is based on the invariant integral, the representation is unitary. Using the exponential map we define the operator

$$dR(X) = \frac{d}{dt}\left\{R[\exp(tX)]\right\}_{t=0} , \quad \text{for all } X \in su(2) . \tag{A.9}$$

This equation introduces a representation of the algebra su(2) on $L^2(\mathrm{SU}(2))$. Since R($g$) is unitary and $T_a^\dagger = T_a$ the operators

$$\mathrm{E}_a = \mathrm{dR}(T_a) \,, \quad a = 1, 2, 3 \tag{A.10}$$

are hermitean. They are the infinitesimal generators of the transformation (A.8). Using (A.9) it follows that

$$\mathrm{R}(g) = \exp(\mathrm{i}X_a \mathrm{E}_a) \quad \text{if} \quad g = \exp(\mathrm{i}X_a T_a) \,, \tag{A.11}$$

and one deduces the commutation relations

$$[\mathrm{E}_a, \mathrm{E}_b] = \mathrm{i}\varepsilon_{abc}\mathrm{E}_c \,. \tag{A.12}$$

The operator

$$\mathrm{E}^2 \stackrel{\text{def}}{=} \sum_{a=1}^{3} \mathrm{E}_a \mathrm{E}_a \tag{A.13}$$

commutes with all generators $\mathrm{E}_a$ and is hence identified with the quadratic Casimir operator of the gauge group SU(2). After choosing a parameterization of the group manifold the infinitesimal generators $\mathrm{E}_a$ act as differential operators in the group parameters on the wave functions $f \in L^2(\mathrm{SU}(2))$. In particular, $\mathrm{E}^2$ is equal to the Laplace-Beltrami operator up to some constant.

The eigenfunctions of the Casimir operator (A.13) transform irreducibly under the gauge group. Thus, to diagonalize $\mathrm{E}^2$ we have to decompose the Hilbert space $L^2(\mathrm{SU}(2))$ into irreducible subspaces. It is well-known that the irreducible representations of SU(2) can be labeled by an integral or half-integral number $j$, the angular momentum. For a given $j$ the representation space has dimension $n = 2j + 1$. So there is a one-to-one correspondence between the angular momentum and the dimensionality of the representation. Within the context of strong coupling perturbation theory it turns out to be more convenient if the irreducible representations are labeled by their dimension. Therefore I will adopt this point of view throughout my thesis.

By means of the Peter-Weyl theorem one determines the representation functions

$$D_{ab}^{(n)}(U) \,, \quad a, b = 1, 2, \ldots, n \,. \tag{A.14}$$

For any fixed value of the indices $n$ and $(a, b)$ they are elements of the Hilbert space $L^2(\mathrm{SU}(2))$. Fortunately we do not need these complicated functions explicitly. What is important is that they span an irreducible subspace of $L^2(\mathrm{SU}(2))$ for a definite value of $n$. Furthermore the Peter-Weyl theorem asserts that there are no further irreducible subspaces in the Hilbert space $L^2(\mathrm{SU}(2))$. It is easy to show that the representation functions $D_{ab}^{(n)}(U)$ are eigenfunctions of the Casimir operator with eigenvalues $(1/4)(n^2 - 1)$.

## A.2 Clebsch-Gordon series

The irreducible representations of the gauge group SU(2) are classified by the quantum number $n$. On $L^2(\mathrm{SU}(2))$ the representation functions (A.14) form a basis of the associated irreducible subspaces. We introduce a simplified notation

$$U_{ab}^n \stackrel{\mathrm{def}}{=} D_{ab}^{(n)}(U) . \tag{A.15}$$

It is often convenient to interpret (A.15) as complex $n \times n$ matrices. Special cases are the trivial representation $n = 1$ where the function (A.15) is just equal to 1, the fundamental representation $n = 2$ where (A.15) are SU(2) matrices

$$U_{ab} \stackrel{\mathrm{def}}{=} U_{ab}^2 , \tag{A.16}$$

and the adjoint representation $n = 3$ where the representation functions can be identified with rotation matrices SO(3). By $\bar{n}$ we denote the irreducible representation that is complex conjugate to $n$

$$U_{ab}^{\bar{n}} = U_{ab}^{n\,*} = U_{ba}^{n\,\dagger} . \tag{A.17}$$

It is a specialty of the group SU(2) that $n$ and $\bar{n}$ are equivalent representations, i. e. all matrices $U^n$ and $U^{\bar{n}}$ are related by the same similarity transformation

$$U^{\bar{n}} = C\, U^n\, C^{-1} . \tag{A.18}$$

For the case of the fundamental representation one finds explicitly

$$U_{ab}^{\bar{2}} = U_{ba}{}^\dagger = \varepsilon_{aa'}\, U_{a'b'}\, \varepsilon_{bb'} , \tag{A.19}$$

where $\varepsilon_{ab}$ denotes the antisymmetric symbol on two indices, with $\varepsilon_{12} = +1$. As a consequence the trace of SU(2) matrices is real. A special case is the adjoint representation. Here 3 and $\bar{3}$ are identical representations.

From a well-known theorem in group theory it follows that the direct product of two irreducible representations $n_1$ and $n_2$ of SU(2) forms a unitary representation of dimension $n_1 \cdot n_2$ of the same group. A basis for the product representation $n_1 \otimes n_2$ on the linear space $L^2(\mathrm{SU}(2))$ is provided by the set of functions

$$U_{a_1 b_1}^{n_1}\, U_{a_2 b_2}^{n_2} , \quad a_1, b_1 = 1, 2, \ldots, n_1 , \quad a_2, b_2 = 1, 2, \ldots, n_2 . \tag{A.20}$$

In general, the product representation is reducible. The decomposition of $n_1 \otimes n_2$ into irreducible components is described by the Clebsch-Gordon series

$$n_1 \otimes n_2 = (|n_1 - n_2| + 1) \oplus (|n_1 - n_2| + 3) \oplus \ldots (n_1 + n_2 - 3) \oplus (n_1 + n_2 - 1) . \tag{A.21}$$

This is just an alternative expression of the well-known result that the addition of two angular momenta with eigenvalues $j_1$ and $j_2$ gives all $j$ values spaced integrally between $|j_1 - j_2|$ and $(j_1 + j_2)$. Corresponding to (A.21) there is a decomposition of the basis functions (A.20) of the product representation

$$U^{n_1}_{a_1 b_1} U^{n_2}_{a_2 b_2} = \sum_{n,a,b} \langle n_1, a_1; n_2, a_2 \mid n, a \rangle U^n_{ab} \langle n, b \mid n_1, b_1; n_2, b_2 \rangle . \tag{A.22}$$

Here the sum extends over all values $n$ that are contained in $n_1 \otimes n_2$ via (A.21). The Clebsch-Gordon coefficients $\langle n_1, a_1; n_2, a_2 \mid n, a \rangle$ can be considered as a $n_1 \cdot n_2 \times n_1 \cdot n_2$ matrix whose first index is the pair $(n_1 n_2)$ and second index the pair $(na)$. By convention this matrix is unitary

$$\sum_{a_1, a_2} \langle n, a \mid n_1, a_1; n_2, a_2 \rangle \langle n_1, a_1; n_2, a_2 \mid n', a' \rangle = \delta_{nn'} \delta_{aa'} , \tag{A.23}$$

$$\sum_{n,a} \langle n_1, a_1; n_2, a_2 \mid n, a \rangle \langle n, a \mid n_1, a'_1; n_2, a'_2 \rangle = \delta_{a_1 a'_1} \delta_{a_2 a'_2} . \tag{A.24}$$

Throughout this thesis we only need some of the Clebsch-Gordon coefficients explicitly. They can be determined by means of a general theorem which states that the product of an irreducible representation and its complex conjugate contains the trivial representation exactly once and the adjoint representation at least once in its Clebsch-Gordon series (cf. ref. [21, sect. 16.6], for example). For the case of SU(2) the associated coefficients are given by

$$\langle \bar{n}, a; n, b \mid 1 \rangle = \frac{1}{\sqrt{n}} \delta_{ab}$$

and

$$\langle \bar{2}, a; 2, b \mid 3, j \rangle = \frac{1}{\sqrt{2}} (\sigma_j)_{ba} , \qquad \langle 3, a; 3, b \mid 3, j \rangle = \frac{1}{\sqrt{2}} \epsilon_{jab} , \qquad \text{etc.}$$

Making use of the equivalence relation (A.19) for the fundamental representations $\bar{2}$ and $2$ one deduces

$$\langle 2, a; 2, b \mid 1 \rangle = \langle \bar{2}, a; \bar{2}, b \mid 1 \rangle = \frac{1}{\sqrt{2}} \varepsilon_{ba} ,$$

$$\langle 2, a; 2, b \mid 3, j \rangle = \frac{1}{\sqrt{2}} (\sigma_j)_{bc} \varepsilon_{ca} ,$$

$$\langle \bar{2}, a; \bar{2}, b \mid 3, j \rangle = \frac{1}{\sqrt{2}} \varepsilon_{bc} (\sigma_j)_{ca} .$$

### A.3  Strong coupling matrix elements

The matrix elements in strong coupling perturbation are such that a product of several link operators corresponding to irreducible representation functions is matched between the static vacuum

$$\langle \Omega \mid U^{n_1}_{a_1 b_1} U^{n_2}_{a_2 b_2} \ldots U^{n_k}_{a_k b_k} \mid \Omega \rangle . \tag{A.25}$$

These objects are zero unless the decomposition of the product contains the trivial representation. The actual value of (A.25) is calculated straightforwardly using the Clebsch-Gordon series (A.21) and (A.22). In the following I give a list of those matrix elements that contribute to the first and second order processes relevant throughout this thesis.

The simplest case is the product of two fundamental representations

$$2 \otimes 2 = 1 \oplus 3 \, ,$$

and the associated non-vanishing matrix element is

$$\langle \Omega \mid \mathrm{U}_{ab} \mathrm{U}_{a'b'} \mid \Omega \rangle = (1/2) \, \varepsilon_{aa'} \, \varepsilon_{bb'} \, . \tag{A.26}$$

In case that the complex conjugate representation $\bar{2}$ occurs instead, the corresponding matrix element can be easily deduced from (A.26) by means of the similarity transformation (A.19), for example

$$\langle \Omega \mid \mathrm{U}_{ab} \mathrm{U}_{a'b'}^{\dagger} \mid \Omega \rangle = (1/2) \, \delta_{ab'} \, \delta_{ba'} \, . \tag{A.27}$$

Three times the fundamental representation does not contain the trivial one, so there are no contributions from such matrix elements. In second order processes we encounter the product of four representation operators. The Clebsch-Gordon series tells us that the intermediate state may be in the trivial or adjoint representation

$$2 \otimes 2 \otimes 2 \otimes 2 = (1 \oplus 3) \otimes (1 \oplus 3) = 1 \oplus 1 \oplus 3 \oplus 3 \oplus 3 \oplus 5 \, .$$

Let P(1) and P(3) denote projection operators on the corresponding irreducible subspaces. The associated matrix elements are

$$\langle \Omega \mid \mathrm{U}_{ab} \, \mathrm{U}_{cd} \, \mathrm{P}(1) \, \mathrm{U}_{a'b'} \, \mathrm{U}_{c'd'} \mid \Omega \rangle = (1/4) \, \varepsilon_{ac} \, \varepsilon_{bd} \, \varepsilon_{a'c'} \, \varepsilon_{b'd'} \, ,$$

$$\langle \Omega \mid \mathrm{U}_{ab} \, \mathrm{U}_{cd} \, \mathrm{P}(3) \, \mathrm{U}_{a'b'} \, \mathrm{U}_{c'd'} \mid \Omega \rangle = (1/12) \, (2 \, \varepsilon_{ac'} \, \varepsilon_{a'c} - \varepsilon_{ac} \, \varepsilon_{a'c'})$$
$$\times (2 \, \varepsilon_{bd'} \, \varepsilon_{b'd} - \varepsilon_{bd} \, \varepsilon_{b'd'}) \, .$$

It is only for the investigation of the seven-link one-particle subspace that more complicated contributions occur. In particular one needs

$$\langle \Omega \mid \mathrm{U}_{ab} \, \mathrm{U}_{cd}^3 \, \mathrm{P}(2) \, \mathrm{U}_{a'b'} \, \mathrm{U}_{c'd'}^3 \mid \Omega \rangle = (1/6) \, \delta_{aa'} \, \delta_{bb'} \, \delta_{cc'} \, \delta_{dd'} \, ,$$

$$\langle \Omega \mid \mathrm{U}_{ab} \, \mathrm{U}_{cd}^3 \, \mathrm{P}(4) \, \mathrm{U}_{a'b'} \, \mathrm{U}_{c'd'}^3 \mid \Omega \rangle = (1/12) \, \varepsilon_{a\alpha} \, \varepsilon_{b\beta} \, \varepsilon_{cc'\mu} \, \varepsilon_{dd'\nu} \, (\sigma_\mu)_{a'\alpha} \, (\sigma_\nu)_{\beta b'} \, .$$

# Appendix B

The Green function (2.70) determines the coordinate dependence of the scattering solution describing the collision of two SU(2) glueballs in two dimensions. In order to calculate the corresponding transition matrix it has to be evaluated at various coordinate points. Substituting the second order approximation of the kinetic energy (3.1) one obtains the integral representation

$$G_{\rm sc}(\boldsymbol{q};\boldsymbol{r}) = (21/2) \int_{\mathcal{B}} \frac{{\rm d}^2 p}{(2\pi)^2} \frac{\exp({\rm i}\boldsymbol{p}\boldsymbol{r})}{\hat{q}^2 - \hat{p}^2 + {\rm i}\rho} \ . \tag{B.1}$$

Mind that this is the two-dimensional counterpart of [1, eq. C.3], so we can adopt many arguments from appendix C of [1]. But cautiously, the dimension is different.

Clearly, there is a representation in terms of Bessel functions

$$G_{\rm sc}(\boldsymbol{q},\boldsymbol{r}) = -(21/4)\,{\rm i} \int_0^\infty {\rm d}t\, F_{\rm sc}(t) \ , \tag{B.2}$$

with

$$F_{\rm sc}(t) = \exp\left[({\rm i}/2)\left(\hat{q}^2 - 4\right) t\right] {\rm e}^{-\rho t} \prod_{k=1}^2 {\rm i}^{r_k} J_{r_k}(t) \ . \tag{B.3}$$

Be careful not to interchange the $t$-integration and the limit $\rho \to 0$ here because the asymptotic behaviour of a product of two Bessel functions is $t^{-1}$. In detail, the asymptotic expansion reads

$$F_{\rm sc}(t) = \sum_{\mu=0}^2 \sum_{\nu=0}^\infty b_{\mu\nu}(\boldsymbol{r}) \exp[-(z_\mu + \rho)t]\, t^{-\nu-1} \ , \quad z_\mu = ({\rm i}/2)\left(8 - 4\mu - \hat{q}^2\right) \ . \tag{B.4}$$

The series that is truncated after $\nu = N$ is denoted by $F_{\rm sc}^N(t)$. As described in detail for the three-dimensional case we estimate the rest (cf. [1, eq. (C.11)]) and obtain an approximation for the Green function with arbitrarily small error $\epsilon$

$$G_{\rm sc}(\boldsymbol{q};\boldsymbol{r}) = -(21/4)\,{\rm i} \left[\int_0^{t_N} {\rm d}t\, F_{\rm sc}(t) + \int_{t_N}^\infty {\rm d}t\, F_{\rm sc}^N(t)\right] + \mathcal{O}(\epsilon) \ . \tag{B.5}$$

To be explicit I demand $\epsilon = 10^{-12}$ and choose $N = 20$. Then the limit of integration turns out to be $t_N = 17$ for $\boldsymbol{r} = 0$, for example.

As far as the evaluation of the finite integral is concerned we can forget about the limit $\rho \to 0$ in (B.3), and the reader is referred to appendix C of [1] once more. We turn to the infinite integral in eq. (B.5). Unless $\nu = 0$ the integration over the summands of the asymptotic expansion (B.4) is well-defined even without the convergence factor ${\rm e}^{-\rho t}$, and we obtain a formula similar to [1, eq. C.13]. So it is only the contributions $\propto t^{-1}$ from the asymptotic expansion of the integrand that make it necessary to discuss the calculation of the Green function in two space dimensions separately. By means of ref. [22, sections 3.381, 8.214] one gets

$$\lim_{\rho \searrow 0} \sum_{\mu=0}^2 b_{\mu 0}(\boldsymbol{r}) \int_{t_N}^\infty {\rm d}t\, \exp\left[-(z_\mu + \rho)t\right] t^{-1}$$

$$= -\lim_{\rho \searrow 0} \sum_{\mu=0}^2 b_{\mu 0}(\boldsymbol{r}) Ei\left[-(z_\mu + \rho)t_N\right] = -\sum_{\mu=0}^2 b_{\mu 0}(\boldsymbol{r}) \left[\ln(z_\mu t_N) + C + \sum_{k=1}^\infty \frac{(-z_\mu t_N)^k}{k \cdot k!}\right] \ ,$$

where $C = 0.577...$ is Euler's constant. For computational purposes the infinite sum is truncated such that the error is less than $\epsilon$.

The above result shows that the Green function (B.1) is logarithmically divergent if $z_\mu = 0$, that is for $\hat{q}^2 = 0, 4$ and $8$. In particular, one finds the threshold behaviour

$$G_{\text{sc}}(\boldsymbol{q};\boldsymbol{r}) = (21/8) \left[ \frac{\ln(\hat{q})}{2\pi} - \text{i} + b(\boldsymbol{r}) \right] + \mathcal{O}(\hat{q}) , \quad b(\boldsymbol{r}) \in \mathbb{R} . \tag{B.6}$$

Contrary to the Ising model, where the Green function was needed only at $\boldsymbol{r} = 0$ and $\boldsymbol{r} = \hat{1}$ for our purposes, the calculation of the transition matrix (2.87) affords the evaluation of $G_{\text{sc}}(\boldsymbol{q};\boldsymbol{r})$ at a considerably larger number of spatial points (cf. eqs. (2.72) and (2.78)). Although the method described above allows to determine the Green function for any value of the relative momentum $\boldsymbol{q}$ and the lattice vector $\boldsymbol{r}$ it is much more elegant and economic to complete this straightforward calculation by some algebraic considerations [24].

Consider the integral representation (B.1). First of all it is easy to deduce the identity (cf. eq. (2.71))

$$(2/21) \left( -\Delta + \hat{q}^2 \right) G_{\text{sc}}(\boldsymbol{q};\boldsymbol{r}) = \delta(\boldsymbol{r}) . \tag{B.7}$$

By means of a partial integration it is equally straightforward to show that

$$G_{\text{sc}}(\boldsymbol{q};\boldsymbol{r} + \hat{k}) - G_{\text{sc}}(\boldsymbol{q};\boldsymbol{r} - \hat{k}) = r_k H(\boldsymbol{q};\boldsymbol{r}) , \qquad k = 1, 2 , \tag{B.8}$$

where $H(\boldsymbol{q};\boldsymbol{r})$ does not depend on the spatial component $r_k$. Joining (B.7) and (B.8) together and assuming that $\boldsymbol{r} \neq 0$ we can determine the unspecified function

$$H(\boldsymbol{q};\boldsymbol{r}) = \frac{1}{r_1 + r_2} \left[ \left(4 - \hat{q}^2\right) G_{\text{sc}}(\boldsymbol{q};\boldsymbol{r}) - 2 \sum_{k=1}^{2} G_{\text{sc}}(\boldsymbol{q};\boldsymbol{r} - \hat{k}) \right] , \tag{B.9}$$

provided that $r_1 + r_2 \neq 0$. After substituting this result in (B.8) again one is left with a recurrence relation for the Green function

$$G_{\text{sc}}(\boldsymbol{q};\boldsymbol{r} + \hat{k}) = G_{\text{sc}}(\boldsymbol{q};\boldsymbol{r} - \hat{k}) + \frac{r_k}{r_1 + r_2} \left[ \left(4 - \hat{q}^2\right) G_{\text{sc}}(\boldsymbol{q};\boldsymbol{r}) - 2 \sum_{k=1}^{2} G_{\text{sc}}(\boldsymbol{q};\boldsymbol{r} - \hat{k}) \right] . \tag{B.10}$$

Due to the planar symmetry $O(2, \mathbb{Z})$ we can restrict attention to the sector $r_1 \geq r_2 \geq 0$. Within this quadrant eq. (B.10) allows one to express $G_{\text{sc}}(\boldsymbol{q};\boldsymbol{r})$ as a linear combination of the values at the corners of the unit square

$$G_{\text{sc}}(\boldsymbol{q};0), \; G_{\text{sc}}(\boldsymbol{q};\hat{1}), \; G_{\text{sc}}(\boldsymbol{q};\hat{1} + \hat{2}) . \tag{B.11}$$

Furthermore, assuming that $\boldsymbol{r} = 0$ one deduces from (B.7)

$$G_{\text{sc}}(\boldsymbol{q};\hat{1}) = (1/4) \left(4 - \hat{q}^2\right) G_{\text{sc}}(\boldsymbol{q};0) + (21/8) . \tag{B.12}$$

# Appendix C

## C.1 The rotation group in two dimensions

The spatial symmetry group of the continuum $(2+1)$-dimensional gauge theory is the group $O(2)$. In addition to the proper rotations about the $z$ axis it contains reflections in a plane containing the $z$ axis. Therefore this group is non-abelian. The irreducible representations can be labeled by positive integers $l = 0, 1, \ldots, \infty$ which are interpreted as the quantum numbers representing the (spatial) angular momentum. Except for the trivial representation $l = 0$, which is one-dimensional, all others are of dimension two. The space of infinitely differentiable functions defined on the unit circle is an infinite-dimensional homogeneous space for the action of the group $O(2)$. The 'circular harmonics'

$$Y_l^m(\phi) = \begin{cases} 1/\sqrt{2\pi} & \text{for } l=0 \,,\, m=0 \\ \exp(iml\phi)/\sqrt{2\pi} & \text{for } l=1,2,\ldots,\infty \,,\, m=\pm 1 \end{cases} \quad (C.1)$$

span the irreducible subspaces which we denote by $\Upsilon_l$.

On a two-dimensional spatial lattice the symmetry group is reduced to the planar rotations $O(2, \mathbb{Z})$. This point group is the symmetry group of the square with undirected sides and is often called the dihedral group $D_4$ (cf. ref. [21], for instance). It is a discrete group of order 8 with 5 irreducible representations $A_1, A_2, B_1, B_2$ and $E$ of dimensionality $1, 1, 1, 1$ and $2$. Through the obvious restriction the irreducible representations $l$ of the continuous group $O(2)$ are also representations of the point group $O(2, \mathbb{Z})$. The corresponding representation functions are the same modulo 4, i.e.

$$D_{m'm}^{(l+4n)}(R) = D_{m'm}^{(l)}(R) \quad \text{for all } n \in \mathbb{N},\ R \in O(2, \mathbb{Z}).$$

The decomposition of each of these representations in terms of the irreducible representations of $O(2, \mathbb{Z})$ is

$$\left.\begin{array}{rcl} 0 &=& A_1 \\ 1 &=& E \\ 2 &=& B_1 \oplus B_2 \\ 3 &=& E \\ 4 &=& A_1 \oplus A_2 \end{array}\right\} \text{ modulo 4 .} \quad (C.2)$$

For lattice calculations it is useful to introduce new basis functions for the subspaces $\Upsilon_l$ which make the above reduction manifest. In the context of two-glueball scattering we can restrict ourselves to even parity solutions, i.e. even $l$ values, because the wave functions are symmetric. As a consequence the $E$-sector, which only contributes to odd subspaces $\Upsilon_l$, is irrelevant. We define (cf. ref. [9])

$$\begin{aligned}
X_{A_1}^0(\phi) &= 1/\sqrt{2\pi} \\
\left. \begin{aligned} X_{A_1}^l(\phi) &= \cos(l\phi)/\sqrt{\pi} \\ X_{A_2}^l(\phi) &= -\mathrm{i}\sin(l\phi)/\sqrt{\pi} \end{aligned} \right\} &\quad \text{for } l = 4, 8, 12, \ldots \\
\left. \begin{aligned} X_{B_1}^l(\phi) &= \cos(l\phi)/\sqrt{\pi} \\ X_{B_2}^l(\phi) &= -\mathrm{i}\sin(l\phi)/\sqrt{\pi} \end{aligned} \right\} &\quad \text{for } l = 2, 6, 10, \ldots
\end{aligned} \qquad (\mathrm{C.3})$$

These functions are a basis for the space of symmetric, infinitely differentiable functions on a unit circle

$$\sum_\Gamma \sum_l X_\Gamma^l(\phi') X_\Gamma^l(\phi)^* = \delta_\mathrm{s}(\phi' - \phi) \,,$$

where the sums run over the representations $\Gamma = A_1, A_2, B_1, B_2$ and the associated angular momentum values. Furthermore, they are orthonormal

$$\int \mathrm{d}\phi \, X_{\Gamma'}^{l'}(\phi) \, X_\Gamma^l(\phi)^* = \delta_{\Gamma'\Gamma} \, \delta_{l'l} \,.$$

## C.2 A simple quantum mechanical example

One or the other of the readers may be not so familiar with quantum mechanical potential scattering in two space dimensions. Those are invited to go through a simple example which reveals some of the kinematical peculiarities.

Consider the relative motion of two identical, non-relativistic bosons of mass $m$ interacting via a rotationally invariant potential

$$\mathrm{V}(r) = \begin{cases} -V_0 < 0 & \text{for } r < R \\ 0 & \text{for } r > R \end{cases} \qquad (\mathrm{C.4})$$

Analogous the well-known three-dimensional case (cf. ref. [23], for instance) the partial wave expansion for the wave functions reads

$$\psi(\boldsymbol{r}) = \sum_{lm} \psi_{lm}(r) Y_l^m(\phi) \,. \qquad (\mathrm{C.5})$$

The radial components can be written as

$$\psi_{lm}(r) = c_{lm}\, u_l(q;r) \,, \qquad (\mathrm{C.6})$$

where $c_{lm}$ is a suitable normalization factor and $u_l(q;r)$ solves the radial Schrödinger equation. The latter has two linearly independent solution given by the Bessel functions $J_l(r)$ and the Neumann functions $N_l(r)$, for example.

The physical wave functions are uniquely determined by means of the regularity condition

$$\lim_{r \to 0} r^{-l}\, u_l(q;r) = 1 \,. \qquad (\mathrm{C.7})$$

In case there is no interaction ($V_0 = 0$) this condition selects the Bessel functions as the regular solutions, that is the wave functions describing the independent motion of the two particles. For large $r$, outside the interaction range, the scattering solutions must behave like the free ones, hence

$$u_l(q;r) \underset{r\to\infty}{\sim} C_l(q) \frac{\cos\left[qr - (l/2)\pi - \pi/4 + \delta_l(q)\right]}{\sqrt{r}} . \tag{C.8}$$

Here $C_l(q)$ is some coefficient, and the phase shifts $\delta_l(q)$ reflect the effect of the potential compared to the free solution.

Now our demonstrative problem can be solved easily. Outside the range of the potential the regular solution is a linear combination of Bessel and Neumann functions

$$u_l(q;r) = \alpha_l(q) J_l(qr) + \beta_l(q) N_l(qr) \qquad \text{for } r > R . \tag{C.9}$$

Substituting the asymptotic expansion for $J_l(qr)$ and $N_l(qr)$ (see ref. [22, section 8.451]) and comparing with (C.8) one obtains a relation between the coefficients $\alpha_l(q)$ and $\beta_l(q)$ and the phase shifts $\delta_l(q)$ of the system

$$\exp\left[2\mathrm{i}\,\delta_l(q)\right] = \frac{\alpha_l(q) - \mathrm{i}\,\beta_l(q)}{\alpha_l(q) + \mathrm{i}\,\beta_l(q)} . \tag{C.10}$$

Inside the interaction region the scattering solution is

$$u_l(q;r) = \gamma_l(q) J_l(\lambda r) \qquad \text{for } r < R \tag{C.11}$$

with $\lambda^2 = q^2 + mV_0$. Demanding continuity for the wave function and its first derivative at $r = R$, the unknown coefficients in (C.9) can be expressed in terms of (C.11). Substituting these results in (C.10) the eigenvalues of the scattering matrix are found to be

$$\exp\left[2\mathrm{i}\,\delta_l(q)\right] = -\frac{q\,J_l(\lambda R)\,H_l^{(2)\prime}(qR) - \lambda\,J_l'(\lambda R)\,H_l^{(2)}(qR)}{q\,J_l(\lambda R)\,H_l^{(1)\prime}(qR) - \lambda\,J_l'(\lambda R)\,H_l^{(1)}(qR)} , \tag{C.12}$$

where we have introduce the Hankel functions

$$H_l^{(1)}(r) = J_l(r) + \mathrm{i}\,N_l(r) , \quad H_l^{(2)}(r) = J_l(r) - \mathrm{i}\,N_l(r) . \tag{C.13}$$

The prime denotes the derivative with respect to $r$.

We analyse the threshold behaviour. For small relative momentum one can replace $\lambda$ by $\lambda_0 = \sqrt{mV_0}$ and make use of the series representations for the Bessel and Neumann functions (cf. ref. [22, sections 8.440, 8.403])

$$J_l(z) = \frac{z^l}{2^l\,\Gamma(l+1)} + \mathcal{O}(z^{l+2}) ,$$

$$\pi\,N_0(z) = 2J_0(z)\left[\ln(z/2) + C\right] + \mathcal{O}(z^2) ,$$

$$\pi\,N_l(z) = 2J_l(z)\left[\ln(z/2) + C\right] - \sum_{k=0}^{l/2} \frac{(l-k-1)!}{2^{2k-l}k!} z^{2k-l} + \mathcal{O}(z^2) \qquad \text{for } l > 0 .$$

Substituting in (C.12) the scattering phases become

$$\delta_0(q) = \frac{2\pi \ln(R_0\, q)}{4\ln^2(R_0\, q) + \pi^2} + \mathcal{O}(q)$$

with

$$R_0 = \exp\left[C + \frac{J_0(\lambda_0 R)}{\lambda_0 R J_1(\lambda_0 R)}\right] \cdot \frac{R}{2}\ ,$$

and

$$\delta_l(q) = \frac{\pi\,(R\,q)^{2l}}{2^l(l-1)!\,\Gamma(l+1)} \cdot \frac{l\,J_l(\lambda_0 R) - \lambda_0\, R\, J_l'(\lambda_0 R)}{l\,J_l(\lambda_0 R) + \lambda_0\, R\, J_l'(\lambda_0 R)} + \mathcal{O}(q^{2l+2}) \qquad \text{for } l > 0\ .$$

As for the well-known three-dimensional case, where the threshold behaviour of the phase shifts is $\delta_l(q) \propto q^{2l+1}$, s-wave scattering is dominant at low energies. So it is only the $l = 0$ channel that contributes to the scattering amplitude near threshold

$$\mathbb{T}(q; \phi', \phi) = \frac{-m^{-1}}{2\pi\,\ln(R_0\, q) - i\pi^2} + \mathcal{O}(q)\ . \tag{C.14}$$

For geometrical reasons the cross section in two spatial dimensions receives a factor $q^{-1}$ (see ref. [20, section 8], for example)

$$\frac{d\sigma}{d\phi'} = \frac{(2\pi)^3}{q}\,\frac{m^2}{4}\,|\mathbb{T}(q; \phi', \phi)|^2\ .$$

Substituting (C.14) the threshold behaviour of the total cross section is found to be

$$\sigma = \frac{4\pi^2}{q}\,\frac{1}{4\ln^2(R_0\, q) + \pi^2} + \mathcal{O}(q)\ .$$